\documentclass[]{aa}
\pdfoutput=1
\usepackage{natbib}
\usepackage{graphicx}
\usepackage{epstopdf} 
\usepackage{txfonts}

\newcommand{\ABAB}{{\tt SimL}}
\newcommand{\FBFC}{{\tt SimS}}
\newcommand{\ABABnosf}{{\tt SimL$^\mathtt{nosf}$}}
\newcommand{\msol}{M$_{\sun}$}

\newcommand{\msolpcsq}{M$_{\sun}\ \mbox{pc}^{-2}$}

\newcommand{\Ts}{$T_\mathrm{s}$}
\newcommand{\Tp}{$T_\mathrm{p}$}
\newcommand{\Omegas}{$\Omega_\mathrm{s}$}
\newcommand{\Omegap}{$\Omega_\mathrm{p}$}
\newcommand{\omegaunits}{km$\,$s$^{-1}$\,kpc$^{-1}$}
\newcommand{\ILRp}{ILR$_\mathrm{p}$}
\newcommand{\ILRs}{ILR$_\mathrm{s}$}
\newcommand{\UHRp}{UHR$_\mathrm{p}$}
\newcommand{\UHRs}{UHR$_\mathrm{s}$}
\newcommand{\CRp}{CR$_\mathrm{p}$}
\newcommand{\CRs}{CR$_\mathrm{s}$}
\newcommand{\OLRp}{OLR$_\mathrm{p}$}
\newcommand{\OLRs}{OLR$_\mathrm{s}$}
\begin{document}

\title{How can double-barred galaxies be long-lived?}

\author{Herv\'e Wozniak \inst{1}\fnmsep\thanks{E-mail:
    herve.wozniak@astro.unistra.fr}}

\institute{Observatoire astronomique de
  Strasbourg, Universit\'e de Strasbourg - CNRS UMR 7550, 11 rue de
  l'Universit\'e, F-67000, Strasbourg, France}


\date{Accepted 2014 11 18. Received 2014 11 18 ; in original form 2014 09 16}

\abstract{ Double-barred galaxies account for almost one third of all
  barred galaxies, suggesting that secondary stellar bars, which are embedded in
  large-scale primary bars, are long-lived structures.  However, up to
  now it has been hard to self-consistently simulate a disc galaxy
  that sustains two nested stellar bars for longer than a few rotation
  periods.}
{ The dynamical and physical requirements for long-lived
  triaxiality in the central region of galaxies still need to be
  clarified.  }
{N$-$body/hydrodynamical simulations including star formation recipes
  have been performed.  Their properties (bar lengths, pattern speeds,
  age of stellar population, and gas content) have been compared with
  the most recent observational data in order to prove that they are
  representative of double-barred galaxies, even SB0. Overlaps in
  dynamical resonances and bar modes have been looked for using
  Fourier spectrograms.
}
{Double-barred galaxies have been successfully simulated with lifetimes
  as long as 7~Gyr.  The stellar and gaseous distributions in the
  central regions are time dependent and display many observed
  morphological features (circumnuclear rings, pseudo-bulges,
  triaxial bulges, ovals, etc.) typical of barred galaxies, even
  early-type.  The stellar population of the secondary bar is younger
  on  average than for the primary large-scale bar.  An important
  feature of these simulations is the absence of any resonance overlap
  for several Gyr. In particular, there is no overlap between the
  primary bar inner Lindblad resonance and the secondary bar
  corotation. Therefore, mode coupling cannot sustain the
  secondary bar mode. Star formation is identified here as possibly being
  responsible for bringing energy to the nuclear mode.  Star
  formation is also responsible for limiting the amount of gas in the
  central region which prevents the orbits sustaining the secondary
  bar from being destroyed. Therefore, the secondary bar can dissolve but
  reappear after $\approx 1$~Gyr as the associated wave is persistent
  as long as central star formation is active.  When star formation is
  switched off the dynamical perturbation associated with the secondary
  bar needs several Gyr to fully vanish, although the central
  morphological signature is almost undetectable after 2~Gyr.  }
{Double-bars can be long-lived in numerical simulations with a gaseous
  component, even in the absence of overlap of resonances or mode
  coupling, provided that star formation remains active, even
  moderately, in the central region where  the nuclear bar lies.  }

\keywords{
Galaxies: kinematics and dynamics --
Galaxies: nuclei --
Galaxies: evolution --
Galaxies: spiral --
Galaxies: bulges
}
\maketitle

\section{Introduction}
\label{sec:introduction}

The early idea that any central triaxial massive component embedded in
a stellar disc could be efficient enough to bring gas close to the
nucleus \citep[e.g.][]{1982modg.proc..113K} has for a long time turned
the attention of both observers and theoreticians to the fruitful
framework of ``bars within bars'' \citep{1989Natur.338...45S} preceded
by a few works dealing with triaxial bulges
\citep[e.g.][]{1988MNRAS.233..337L}.  At that time, the bars within
bars scenario was sustained by a few crucial observations
\citep[e.g.][]{1974IAUS...58..335D, 1986ApJS...61..631B}. Since then,
the observation of small samples of double-barred galaxies
\citep[e.g.][]{1993AJ....105.1344B, 1995A&AS..111..115W} laid the
groundwork for larger samples, less biased, that has refined the
global picture by removing from the list of prototypes a few false
detections made in the optical and adding a lot of new examples mainly
detected in the near infrared \citep[][for a comprehensive
  review]{2011MSAIS..18..145E}.

However, the formation of nuclear bars has remained puzzling in spite
of the efforts made by several groups to perform realistic modelling of
these complex systems. Our difficulty in creating  a standard scenario
for the evolution of the central kpc is certainly not due to the lack
of models that show how the stellar material can be assembled
to form the inner bar, but  rather is due to the coexistence of several
credible physical mechanisms. Most of these mechanisms have been
studied mainly through numerical simulations (N$-$body and/or
hydrodynamical simulations) although orbit analysis
\citep{1997ApJ...484L.117M,2000MNRAS.313..745M,2010ApJ...719..622M}
has helped a lot to understand the foundation of double-bar dynamics.

The double-bar formation scenarios can  first be divided into two major
classes: either the large-scale bar forms first, i.e. before the inner
one, or the contrary. The first case is sustained by early theoretical
works \citep{1989Natur.338...45S}, simulations with a gaseous
component present \citep{1993A&A...277...27F, 1994mtia.conf..170C,
  1996A&AS..118..461F}, but also pure collisionless simulations
\citep{1999A&A...348..737R,2006A&A...447..453C} making the need for a
gaseous component questionable if the initial stellar disc is
dynamically hot and embedded in a massive halo.  The second case was
 considered  hypothetical until \citet{2007ApJ...654L.127D}
created such examples by fine tuning the initial conditions of their
collisionless simulations.  One can argue that a third case should
exist, that is the two bars growing simultaneously
\citep{2013MNRAS.433L..44S}. It seems in all cases that the outer bar forms
slightly before the inner bar.

Whatever the formation scenario is, the main concern with inner bars
formed in N$-$body/hydrodynamical simulations is that they are
short-lived because the central gas concentration naturally tends to
destroy the bar. However, the lifetime of any nuclear bar must be long
enough to be compatible with the high frequency of double-bars:
30~\%\ of barred galaxies or 20~\%\ of all galaxies
\citep{2011MSAIS..18..145E}. Past simulations that were able to
produce long-lived double-barred models
\citep[e.g.][]{2002MNRAS.337.1233R,2007ApJ...654L.127D} are either 2D
or purely collisionless.

We show here that 3D N$-$body/hydrodynamical simulations,
implementing classical star formation recipes, are able to create
long-lived double-barred galaxies. We identify the dynamical and
physical processes whereby a double-barred system can survive on
several Gyr.

This paper is focused on the long-term evolution of double-barred
galaxies. In Sect.~\ref{sec:model} we describe our simulations and
introduce postprocessing techniques. The evolution of the central
region is addressed in detail in Sect.~\ref{sec:evol}. As forthcoming
papers dealing with other aspects of double-barred galaxies will be
based on this topic, we give  a detailed description there.  We compare the
properties of simulations with observational constraints in
Sect.~\ref{sec:comparison}. Finally, we discuss our results in
Sect.~\ref{sec:discussion} and conclude in the last section.


\section{Description of the numerical simulations}
\label{sec:model}

\begin{table*}
\caption{Main initial parameters: name of the run (Model), simulation
  length, numbers of stellar ($N_\mathrm{s}$) and gas ($N_\mathrm{g}$)
  particles, masses and scalelengths of the two Miyamoto-Nagai initial
  distributions ($M_1, M_2, l_1, l_2$), mass and scalelength of the gas
  distribution ($M_\mathrm{g}, l_\mathrm{g}$). The last column contains
  the number of stellar particles at the end of the simulation
  ($N_\mathrm{s}^\mathrm{end}$).}
\label{tab:simul}
\flushleft
\begin{tabular}{lrrrcclllrrrr}
\hline \hline
Model &  End& $N_\mathrm{s}$ & $N_\mathrm{_g}$ & $M_1$ & $M_2$ & $l_1$ & $l_2$ & $M_\mathrm{g}$ & $l_\mathrm{g}$ & $N_\mathrm{s}^\mathrm{end}$ & \cr
      &(Myr)&$\times 10^{6}$&$\times 10^{4}$&$\times 10^{11}$ M$_{\sun}$&$\times 10^{11}$ M$_{\sun}$&(kpc)&(kpc)&$\times 10^{11}$ M$_{\sun}$&(kpc)&$\times 10^{6}$    & \cr
\hline
\ABAB & 9486& 2.5 &5 & 0.1 & 1.0 & 1.5 & 6.5 &0.11 & 6.0 & 3.32& \cr
\FBFC & 5798& 2.5 &5 & 0.1 & 1.0 & 1.0 & 3.5 &0.11 & 3.0 & 3.21& \cr
\hline
\end{tabular}
\end{table*}

We  performed several simulations of disc galaxies varying mainly
the initial disc scale lengths and the gas contents.  For each
simulation, the conditions for the initial stellar and gaseous
populations were computed as follows.  The initial stellar
population was set up to reproduce a typical disc galaxy. The positions
and velocities for $N_s$ particles were drawn from a superposition of
two axisymmetrical \citet{MN75} discs of mass respectively $M_1$ and
$M_2$ (cf. Table~\ref{tab:simul}).  The shape of Miyamoto \& Nagai
density distribution depends on the choice of two parameters,
traditionally called $a$ and $b$. The \citet{1911MNRAS..71..460P} sphere
($a=0$) and the \citet{k56} flat disc ($b=0$) are the two extreme
possible distributions. The use of Miyamoto \& Nagai distributions
enabled us to create initial conditions very close to typical disc
galaxies. The real radial scale lengths are respectively $l_1=a_1+b$
and $l_2=a_2+b$. We  chose a scale height of $b=0.5$~kpc common
to all simulations.  Thus, with our choice of parameters
(cf. Table~\ref{tab:simul}), the first component can be viewed as
representing a bulge, the second one a disc, with the main
advantage that there is no discontinuity in either the mass density
distribution or the gravitational forces. The initial velocity
dispersions were computed by solving numerically the Jeans equations. The
gaseous component is represented by $N_g$ particles for a total mass
of $M_g$ distributed in a $l_g$ scalelength Miyamoto-Nagai disc.

Some differences in the initial conditions should be noted.
The disc scalelength of \ABAB\ ($l_2=6.5$~kpc, hence the "L" for
"long") is almost twice that of \FBFC\ (3.5~kpc, "S" for "short") for
the same scaleheight ($b=0.5$~kpc). The dynamical timescales of the
stellar bar development should thus be longer for \ABAB\ than for
\FBFC.

The dynamical evolution was computed with a particle--mesh N-body code
that includes stars, gas, and recipes to simulate star formation. The
broad outlines of the code are the following. The gravitational forces
are computed with a particle--mesh method using a 3D polar grid with
$(N_R, N_\phi, N_Z)=(60,64,312)$ active cells, leading to a vertical
sampling of 50~pc. The smallest radial cell in the central region is
36~pc large.  The hydrodynamics equations are solved using the SPH
technique, following closely the implementation suggested by
\citet{b90}. Since we used a log--polar grid, we  improved the
pre-computation of self-forces by subdividing each cell in $(n_r,
n_\phi, n_z)=(32,6,6)$ subcells. Self-forces were then linearly
interpolated before being subtracted from gravitational forces.  The
spatial and forces resolutions are thus much higher than in our
previous studies based on the same code
\citep[e.g.][]{2006A&A...452...97M,2006MNRAS.369..853W}.

The star formation and feedback modelling is based on the
instantaneous star formation approximation \citep[see][and reference
  therein, for details]{2004A&A...421..863M}. The major steps are 1)
the identification of the regions where star formation can be ignited,
2) the conversion of a fraction of gas into stars, and 3) the
computation of the amount of energy and metals injected in the
interstellar medium (i.e. the energy and chemical feedback from type
II supernovae -- SNII). Because the last step leads to gas
heating, a simple treatment of radiative cooling is implemented
in the energy equation.

The first task is to identify the gaseous particles that will form
stars. \citet{FB93} examined some possible criteria. Not surprisingly,
they found that the standard Jeans instability criterion can be
applied to spherical non-rotating gaseous systems, but  for
rotating flat discs that Toomre's instability criterion
\citep{1964ApJ...139.1217T} is a better indicator. Observational
evidence also appears to support the use of this criterion as a good
indicator for locating star formation at intermediate scalelengths
\citep[][and references therein]{1998ARA&A..36..189K}. In all cases, a
particle $i$ will be assumed to undergo a star formation episode if
the following condition is verified,
$$
Q_{i}^{g} = \frac{s_{i}\kappa_{i}}{\pi G \Sigma_{i}^{g}} \leq \lambda
,$$ where $Q_i^g$ is Toomre's parameter, $s_{i}$ is the local sound
speed, $\kappa_{i}$ is the generalized epicyclic frequency, and
$\Sigma_{i}^{g}$ is the gas surface density.  The constant $\lambda$
equals unity in the case of an axisymmetric gaseous disc subject to
radial instabilities \citep[cf.][]{1964ApJ...139.1217T}. However, a
value of $\lambda \approx 1.4$ (with $s = 6$ km s$^{-1}$) has been
derived from observations \citep{1990ASSL..161..405K}. Since we intend
to reproduce realistic conditions for star formation, we  adopted the same value for $\lambda$.

The pattern speed \Omegap\ of the large-scale bar is easily determined
using its position-angle measured every 10~Myr during the simulation
run. Any post-processing technique applied on snapshots (spaced by
50~Myr) or on movies (timestep of 1~Myr) gives the same results. For
the nuclear or inner bar, the determination of \Omegas\ is 
more difficult. The initial size of the nuclear bar is unpredictable, and so it is
difficult to catch its position-angle in real time during the
simulation run. The snapshot sampling (every 50~Myr) is too long to
make a proper measurement as the rotation period of the nuclear bar can
be much shorter. Therefore, the nuclear position-angle was determined
directly from the movies using the inertia moment technique confirmed by
eye measurements. However, the errors can be large (typically $\approx
2$~Myr on rotation period measurements), especially when the nuclear
bar rotation period $T_\mathrm{s}$ is of the order of a few tens of
Myr.  A Fourier transform technique (as used by
\citealp{2011MNRAS.417..762Q}) is unapplicable here because of the
frequency cutoff. Therefore, for a few periods of interest,
simulations have been partly recomputed using a smaller output
timestep (1~Myr) in order to apply this technique (see
Sect.~\ref{sec:discussion}). For these particular periods, this
technique confirms our manual measurements well within our
conservative error bars.

The terminology used in the case of double-barred galaxies is wide and
thus confusing. We thus clarify hereby what terms will be used
throughout this paper. The simulations will represent double-barred
galaxies, which means two stellar bars will coexist. A \emph{\emph{primary}}
bar is the main bar with respect to its size. It is a large-scale
structure. It is often named \emph{\emph{large-scale}} bar. This does not
imply that it appears first, even if this is the case for our
simulations. A \emph{\emph{secondary}} bar is a smaller stellar structure
embedded in the primary bar. However, it is often called
the \emph{\emph{inner}} or \emph{\emph{nuclear}} bar.  Hereafter, in the context of
this paper, we  use the word \emph{\emph{``nuclear''}} for a rapidly rotating
secondary bar extending less than 500~pc across, and reserve the term
\emph{\emph{``inner''}} for larger secondary bars.

\section{Evolution of the central regions}
\label{sec:evol}

\begin{figure*}
\centering
\includegraphics[keepaspectratio,width=\hsize]{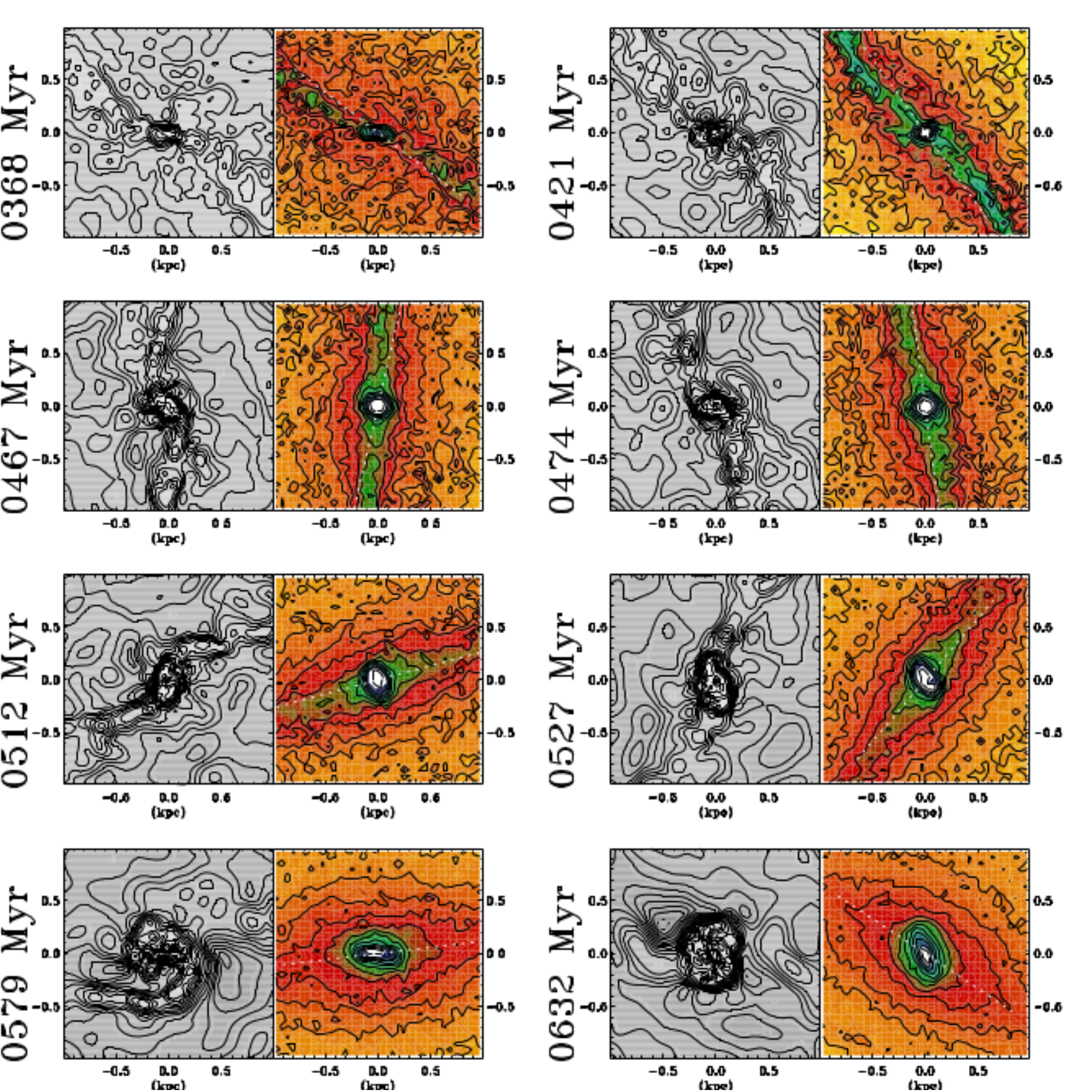}
\caption{Run \FBFC: gas mass surface density (left) and B$-$Cousins
  surface brightness (right) in the central $\pm 1$~kpc for $t <
  0.7$~Gyr. Isocontours are spaced by 0.15 magnitude in $\log
  \mathrm{M}_{\sun}\,\mathrm{pc}^{-2}$ for the gas and 0.5
  $\mathrm{Bmag}\,\mathrm{pc}^{-2}$ for the stars. The map resolution
  is 10~pc. Gas particles have been convolved by the SPH kernel
  to obtain the real distribution. Gas surface densities below
  1~$\mathrm{M}_{\sun}\,\mathrm{pc}^{-2}$ are contoured with dashed
  lines. The straight dotted line is the primary bar position-angle.
}
\label{fig:fbfc}
\end{figure*}

In Figs.~\ref{fig:fbfc2} and \ref{fig:abab} we show the face-on
projections of the central region ($2\times 2$~kpc) for the two runs
used throughout this paper.  Figure~\ref{fig:fbfc} focuses on the
$1\times 1$~kpc of \FBFC\ during the formation of the nuclear bar.
The stellar mass distribution has been photometrically calibrated
using the technique of \citet{2004A&A...421..863M}. A blue photometric
band (B Cousins) has been chosen  to emphasize the regions with
the youngest stellar population. For this purpose, the age of the
initial stellar population has been kept fixed at 10~Gyr for all
snapshots. This implies that a direct photometric comparison of
snapshots has no sense since the initial stellar population is not
getting old and thus its luminosity is not dimming. Because it is a
post-processing technique, this obviously has no impact on the
self-consistent chemodynamical evolution of the simulation.

\begin{figure}
\centering
\includegraphics[keepaspectratio,width=\hsize]{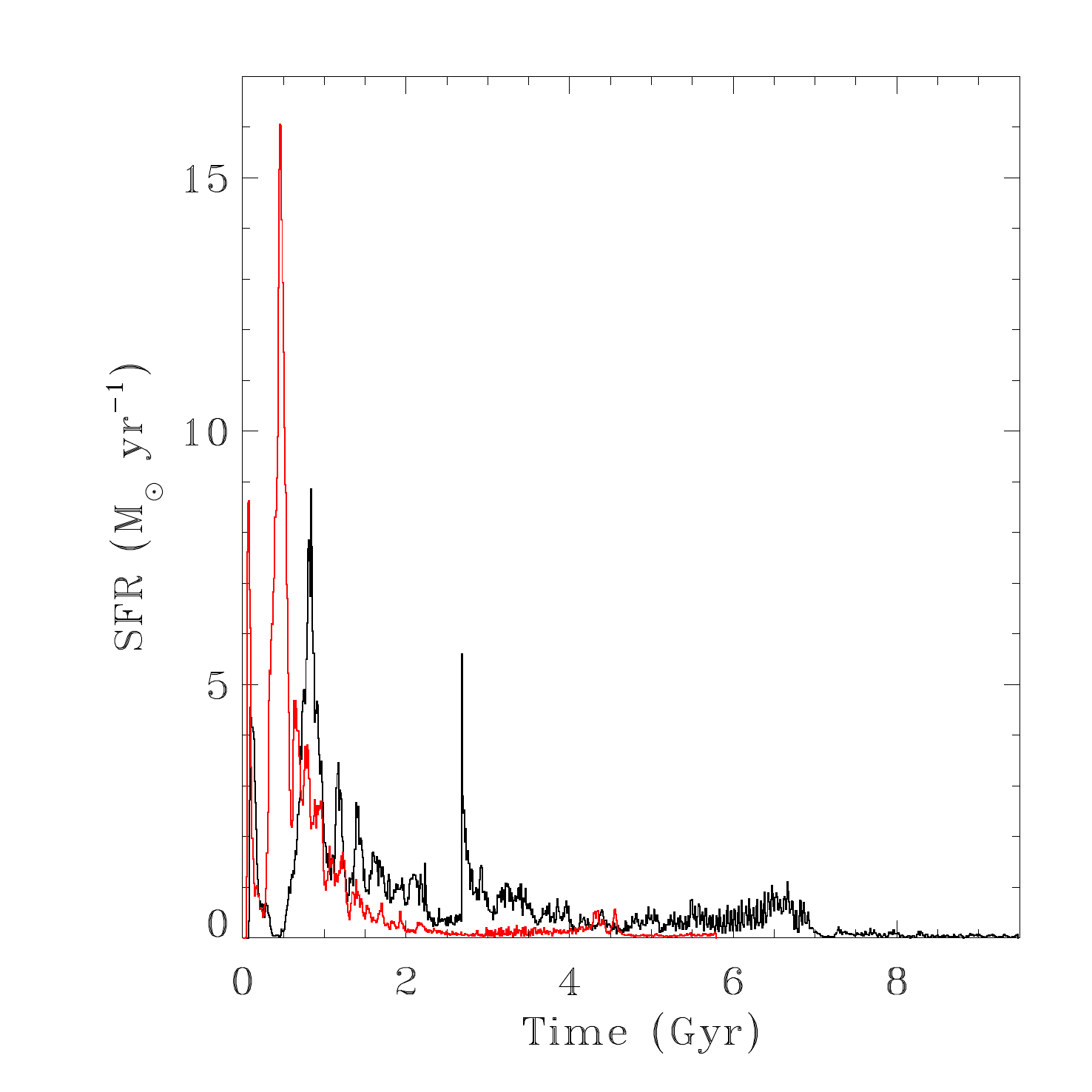}
\caption{Star formation rate (SFR) averaged in a sliding window of
  10~Myr for runs \FBFC\ (red) and \ABAB\ (black).}
\label{fig:SFR}
\end{figure}

The two simulations show different histories of star formation
(Fig.~\ref{fig:SFR}). Their dynamical history is obviously different,
the main driver being the dynamical timescale that is roughly doubled
for \ABAB\ with respect to \FBFC.

%
%
\subsection{Run \FBFC}
\label{ssec:fbfc}

\subsubsection{The nuclear bar formation}
\begin{figure*}
\centering
\includegraphics[keepaspectratio,width=\hsize]{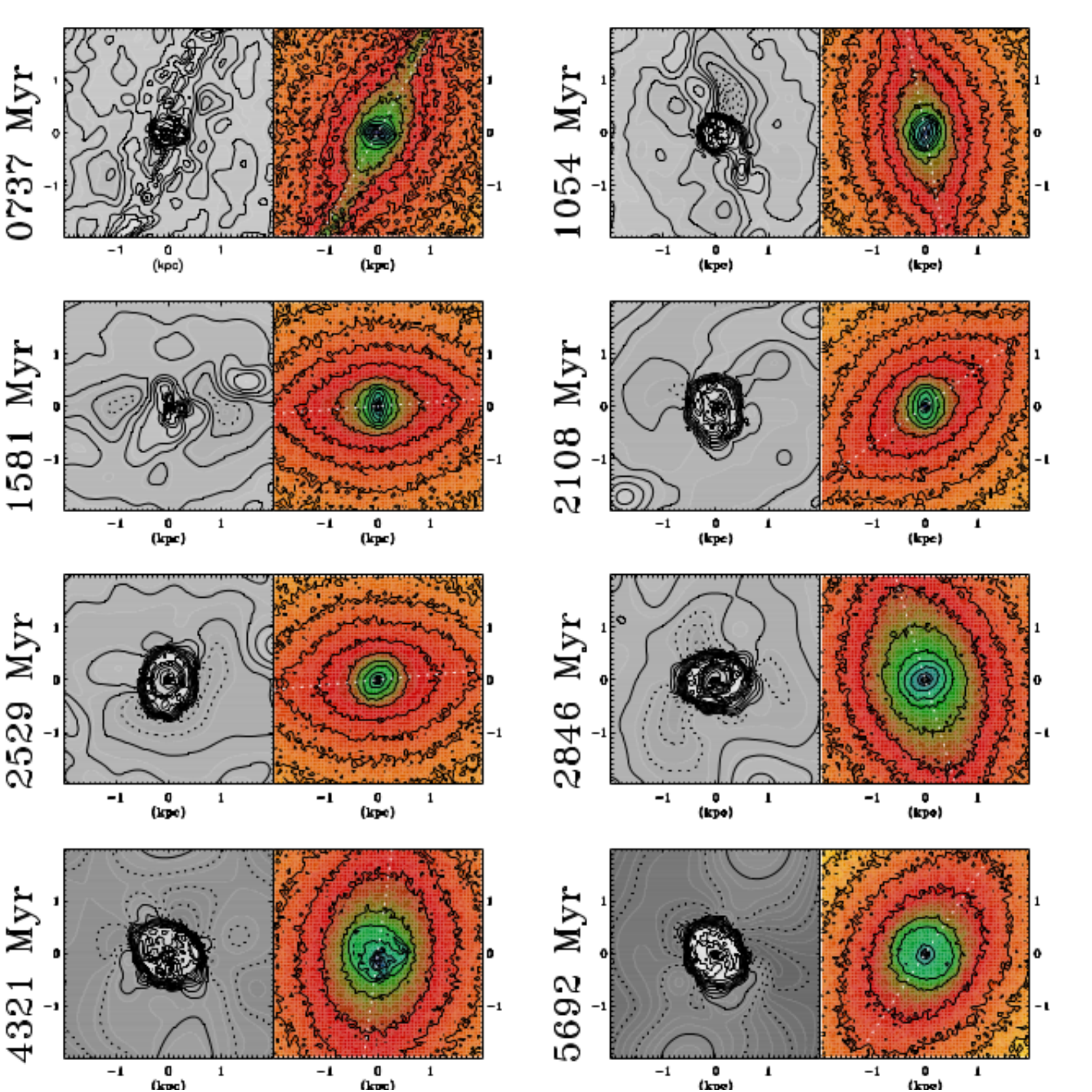}
\caption{As in Fig.~\ref{fig:fbfc}, but for $t > 0.7$~Gyr in the central $\pm
  2$~kpc. The map resolution is now 20~pc. Gas isocontours are spaced
  by 0.25 magnitude in $\log \mathrm{M}_{\sun}\,\mathrm{pc}^{-2}$.
  The radius of \ILRp\ is indicated by dashed white circles.}

\label{fig:fbfc2}
\end{figure*}

The initial disc quickly develops a typical strong bar and a spiral
structure both in the stellar and the gaseous components. The gravity
torques due to the bar and spiral structure drive the gas inwards and
reorganize the mass distribution even for the old stellar population;
this gas inflow occurs in a rather short timescale since the star
formation rate peaks at $t=0.45$~Gyr (cf. Fig.~\ref{fig:SFR}). As
expected, star formation also occurs along gaseous spiral arms that are not
shown in Figs.~\ref{fig:fbfc} and \ref{fig:fbfc2}.

>From $t\approx 0.36$~Gyr, the gas distribution in the central 400~pc
(Fig.~\ref{fig:fbfc2}) starts to twist by an increasing angle that
amounts to $\approx 45^\circ$ at $t=0.4$~Gyr. This twist is also
visible in the distribution of the new stellar population. At the
beginning of the twisting process, the new stellar population is
aligned with the gas distribution. Afterwards, the gas twist angle
still increases, quickly reaching  $90^\circ$ at $t\approx 0.42$~Gyr,
but not the stellar angle since the collisionless component is less
reactive to the torque induced by the local gravitational potential
twist. Moreover, the stellar mass trapped in that region also
increases as a result of the on-going star formation, thus making   the
decoupling of the stellar component easier. When the gas twist angle
continues to increase so that the central gas distribution eventually
realigns with the larger scale flows, the new stellar population is
rather axisymmetric.

At $t\approx 0.5$~Gyr there is, however, as much mass in the new stellar
component as in the gas component.  A second episode of twisting then
begins.  At $t\approx 0.51$~Gyr, the angular motion of the new stellar
population distribution becomes elongated and definitively
decouples from the gas distribution. For $t\approx 0.52$~Gyr, the central gas
twist is again almost perpendicular to the major-axis large-scale
stellar bar. At that point, one can clearly identify a so-called
nuclear secondary bar whose rotating motion seems independent from
the gas motion.

\begin{figure}
\centering
\includegraphics[keepaspectratio,width=\hsize]{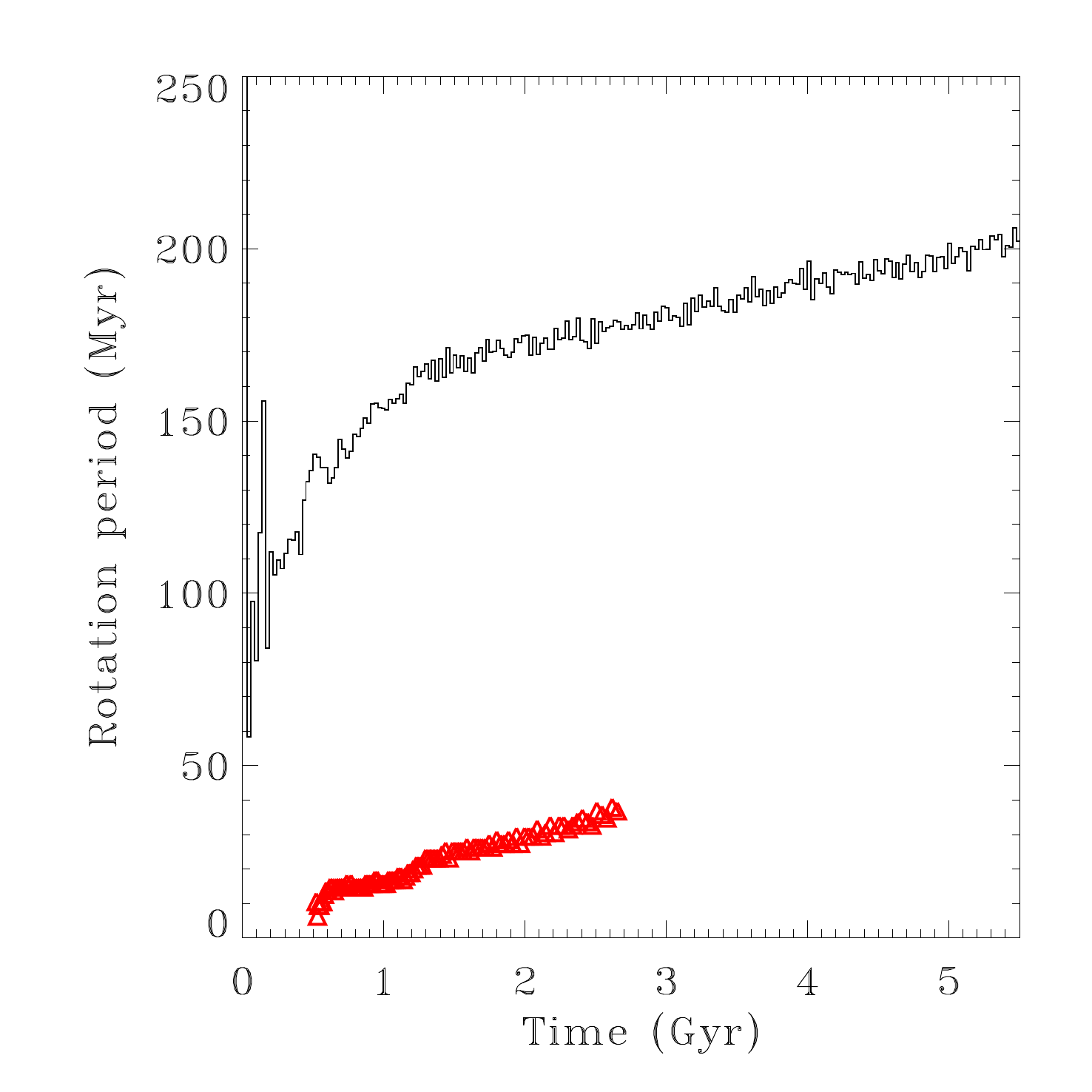}
\caption{Period of rotation for the primary bar (in black) and the
  nuclear bar (in red) for \FBFC. }
\label{fig:FBFC_period}
\end{figure}

During this formation phase, the pattern speed of the nuclear bar is
very high, leading to a low rotation period of $T_\mathrm{s}\approx
10$~Myr (Fig.~\ref{fig:FBFC_period}). This value is the dynamical
signature of the gas from which the new stellar population has been
created. The radius of the nuclear bar is then $l_\mathrm{s} \approx
250$~pc, whereas $l_\mathrm{p}\approx 4.55$~kpc, which leads to
$l_\mathrm{s}/l_\mathrm{p} \approx 0.056$.

\subsubsection{The nuclear bar evolution}
Once the nuclear bar has been formed and rotates independently of
the gas (what was called ``decoupling'' in some studies from the 1990s), the
central gas concentration quickly gets aligned with the stellar
component.  For $t\ga 0.63$~Gyr the gas dynamics can be considered 
fully driven by the nuclear bar evolution (orientation, velocity)
since the gas no longer represents the most massive component in that part of
the galaxy. The morphology is modified consequently. Indeed,
the gas distribution is more ring-like around the nuclear bar or
sometimes disc-like, especially when the energy feedback from
SNII temporarily dissolves a part of the gaseous ring.

After 1~Gyr, the total mass inside the central kpc has increased by a
factor of 1.5. Indeed, a nuclear gas disc is formed from the accumulation
of gas in the centre, and new stars are actively formed there.  The
first cause of this mass inflow is the overall reorganization of the
mass distribution under the influence of the stellar bar, even the old
population. Owing to gravitational torques exerted on the gas by the
stellar bar, the extra mass in the form of gas and new stars amounts
to $1.97\,10^{9}$~M$_{\sun}$ at $t=1$~Gyr, which is only 31\%\ of the
whole additional mass. The redistribution of the old stellar
population contributes to the other 69\%.

The length of the nuclear bar slowly increases as it slows down. The
large-scale bar also increases, but at a lower rate. Therefore, at $t=2$~Gyr
$l_\mathrm{s} \approx 1$~kpc, $l_\mathrm{p} \approx 5$~kpc so that
$l_\mathrm{s}/l_\mathrm{p} \approx 0.2$.

\subsubsection{The nuclear bar dissolution. Formation of the nuclear disc}
While the nuclear stellar bar continuously grows, the gaseous ring
radius increases. For $t \ga 2$~Gyr the nuclear bar gets also thicker,
looking sometimes like an oval surrounded by a ring of gas.  The ring
itself gets broader as the nuclear bar thickens.  The nuclear bar is
sometimes slightly off-centred, in particular around $t\approx
1.6$~Gyr, but this is a transient effect.

It is impossible to precisely date the moment when the nuclear bar
can be considered as dissolved because the dissolution process is smooth
and has a long timescale. Except for $t \ga 2.7-3.0$~Gyr, one can
consider that the nuclear bar has been replaced by a nuclear disc.
This stellar nuclear disc has a clear signature in the edge-on mass
distribution. It also drives the shape of the gaseous nuclear
ring. Any slight temporary ovalization of the stellar nuclear disc
leads to an alignment of the major-axis of the gaseous ring.

Since the gas accumulation in the central region continues, whereas
the star formation rate is very low in the ring, the gas mass
increases in the ring and the nucleus. For $t > 4$~Gyr the gaseous
ring becomes unstable. It starts to twist and collapse. This permits
the gas flows to reach the nucleus and leads the star formation rate
to briefly increase (Fig.~\ref{fig:SFR}). As a result, the nice broad
gaseous ring is replaced by a gaseous disc of the same size as the
nuclear disc for $t > 4.3$~Gyr until the end of the simulation.

\subsection{Run \ABAB}
\label{ssec:abab}
%
%

\begin{figure*}
\centering
\includegraphics[keepaspectratio,width=0.95\hsize]{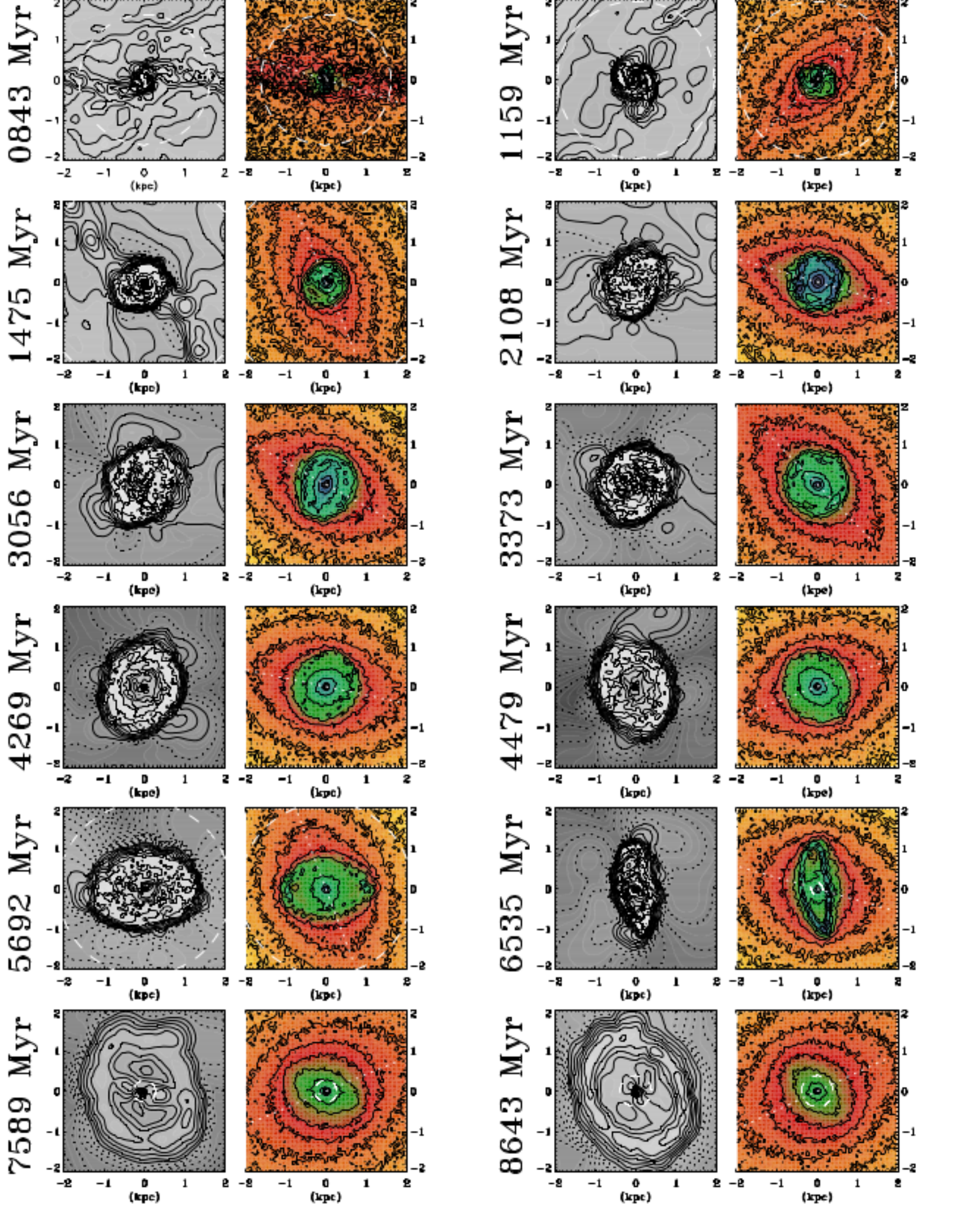}
\caption{As in Fig.~\ref{fig:fbfc2}, but for \ABAB.  From $t=6535$~Myr
  the dashed white circle represents \ILRs\ instead of \ILRp\ which is
  outside the field of view.  }
\label{fig:abab}
\end{figure*}

\subsubsection{The nuclear bar formation: the first secondary bar}
As for \FBFC, the initial disc of \ABAB\ develops a typical large-scale strong
bar and spiral structure. However, because the disc scalelength of
\ABAB\ is roughly twice that of \FBFC, the dynamical timescale for
the development of the main stellar bar is significantly greater. It
is only after $t\approx 0.8$~Gyr that the new stellar population
created in the central gas concentration forms a nuclear bar-like
structure (Fig.~\ref{fig:abab}). This nuclear structure is very small,
$l_\mathrm{s} \approx 250$~pc, whereas $l_\mathrm{p}\approx 5$~kpc,
which leads to $l_\mathrm{s}/l_\mathrm{p} = 0.05$. It is also
transient,  lasting less than $\approx 0.6$~Gyr, and is often asymmetric
and/or off-centred. It then dissolves into a stellar nuclear spiral
structure surrounded by a circumnuclear gaseous ring. It must 
be also mentioned that the large-scale bar growth is not completed yet.
However, even if the length of both bars increases during this phase,
the ratio $l_\mathrm{s}/l_\mathrm{p}$ remains approximately
constant. For instance, at $t\approx 1.3$~Gyr,
$l_\mathrm{s}/l_\mathrm{p}\approx 0.3/6.0 = 0.05$.

\begin{figure}
\centering
\includegraphics[keepaspectratio,width=\hsize]{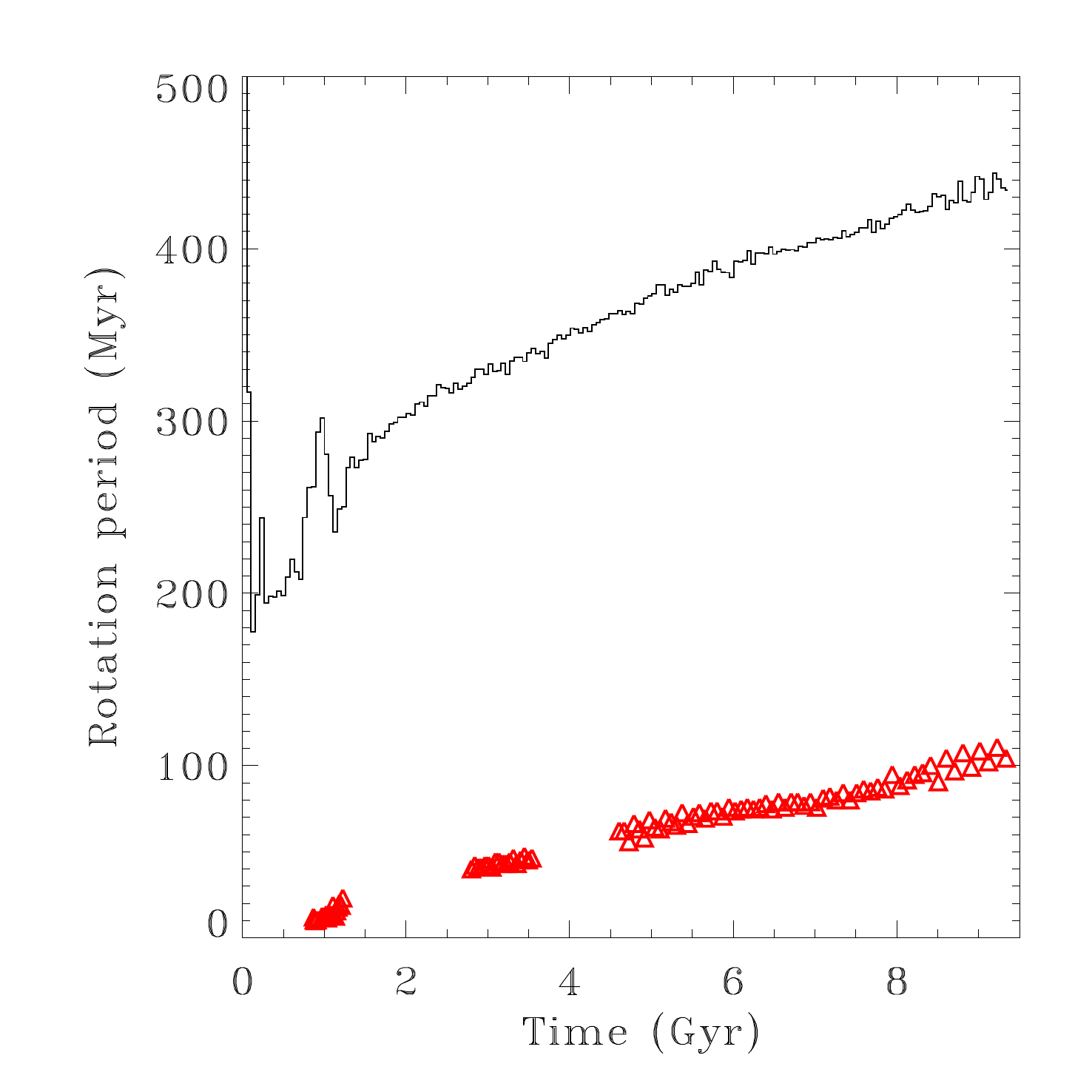}
\caption{As in Fig.~\ref{fig:FBFC_period} for \ABAB.}
\label{fig:ABAB_period}
\end{figure}

Because of its time dependent morphology, the pattern speed of this
nuclear structure is difficult to measure precisely. The pattern
rotation period of the large-scale bar and the nuclear structure are
shown in Fig.~\ref{fig:ABAB_period}. It was not possible to
determine properly the nuclear rotation period for the whole lifetime
because of the quick changes in morphology. However, the nuclear
structure has a very short rotation period (between 10 and 30 Myr)
that strongly contrasts with the primary bar rotation period; it is at least ten times
longer. During the lifetime of the nuclear structure, the large-scale
bar temporarily slows down, the rotation period increasing from
200~Myr to 300~Myr (Fig.~\ref{fig:ABAB_period}).

When the nuclear structure dissolves, both a gaseous disc and a
stellar disc (mainly made of new stellar population), surrounded by a
ring-like structure, survive. These discs form  the remnant of the
nuclear bar. The size of both gaseous and stellar discs increases 
to $\approx 1.8$~kpc in diameter, which leads to
$l_\mathrm{s}/l_\mathrm{p}\approx 0.9/8 =0.11$. However, because the remnant
stellar disc is no longer barred, $l_\mathrm{s}/l_\mathrm{p}$ has
 another meaning here. Transient gaseous and/or stellar spiral arms
often appear inside and outside the ring-like structure until
$t\approx 2.6$~Gyr.

\subsubsection{The inner bar formation: the second secondary bar}

After $t\approx 2.6$~Gyr, the inner region returns to a bar-like
shape. The pattern speed can be determined much more precisely now than
for the first nuclear bar. The gaseous and stellar components both
display the same features. The inner bar ends in a broad circumnuclear
ring. In the centre, the remnant of the first nuclear bar settles in
the centre as a flattened small bulge.  As the stellar inner bar
begins to turn into an oval, the gaseous counterpart progressively
dissolves. Only a gaseous circumnuclear ring survives, that gradually
gets broader (up to roughly 500~pc across the torus).

For $t\ga 4$~Gyr, it becomes difficult to unambiguously determine a
nuclear bar structure. The central kpc region looks like the triaxial
pseudobulges  described by \citet{2004ARA&A..42..603K}.  However,
even if the contrast between the inner structure and the old stellar
population background is too low to permit a proper pattern speed
determination, the inner region is not devoid of
non-axisymmetries. Therefore, for $t \ga 4.5$~Gyr, the whole inner
stellar structure, including the now vanishing circumnuclear ring,
turns into a strong oval that looks like a $l_\mathrm{s}=2.2$~kpc long
bar ($l_\mathrm{s}/l_\mathrm{p}\approx 0.275$). The shape of this
second (in the time sequence) inner bar depends strongly on its
relative orientation with respect to the large-scale bar. The
higher axis ratio is obtained when the angle between the two bars
reaches 90\degr. When the two bars are aligned, the axis ratio is
roughly halved.

For $t\ga 4.5$~Gyr, the responsive gaseous ring also starts  to turn
into an oval, meaning that the gas ring shrinks and expands depending
on the relative orientation of the inner bar with respect to the
large-scale bar. The ring progressively dissolves into a disc that
closely follows the same changes in morphology as the stellar
nuclear bar and thus shares the same orientation.

Because star formation, which mainly occurs in the central region, consumes
the gas, the nuclear disc size significantly decreases for $t \ga
6$~Gyr and becomes dynamically unstable. For $6 < t < 7$~Gyr,
the star formation rate temporarily increases
(Fig.~\ref{fig:SFR}) because of the collapse of the gaseous
nuclear ring. This is the last episode of star formation in that
region. At $t\approx 7$~Gyr, the gaseous nuclear disc has almost fully
disappeared, whereas the stellar counterpart is made of an inner bar
with $l_\mathrm{s}\approx 1.3$~kpc embedded in a slightly larger
stellar disc. The large-scale bar has $l_\mathrm{p}=8.5$~kpc leading
to $l_\mathrm{s}/l_\mathrm{p}\approx 0.15$.

As the stellar inner bar (oval-like) still exists for $t > 7$~Gyr, the
gas inflow continues in such a way that a gaseous circumnuclear ring
forms again. The gas mass surface density in the ring, and in the
region encircled by the ring, is higher than 1~\msolpcsq\ , but never
reaches the same high values as during the first 2~Gyr. In
particular, the gas distribution is smoother.

These structures, the inner stellar bar or oval and gaseous circumnuclear
ring, last until the end of the simulation at $t\approx 9.5$~Gyr
without any significant morphological change. The only noticeable fact
is the permanent slow down of both bars due to the classical angular
momentum exchange with the large-scale disc.

At the end of the simulation, $l_\mathrm{s}/l_\mathrm{p}\approx 1.5/9
= 0.17$, quite close to the value at $t=7$~Gyr. Indeed, the two bars
slow down, but at different rates.

\section{Comparison with observations}
\label{sec:comparison}

\subsection{Gas mass in the central kpc}
\label{ssec:gasmass}

It has been argued in the past \citep[][and references
  therein]{2004ApJ...603..495P} that large amounts of gas are not
required to form and sustain a double-bar. However, if the nuclear bar
is long-lived, and we can imagine that the galaxy has been gas rich in the
past, then its gas has been consumed and possibly redistributed
thus explaining  the current low content. Alternatively, if the nuclear
bar is a short-lived phenomenon, one has to define the lower gas mass
limit to initiate the formation process. 
This means that the expressions `gas rich' or `gas poor' used in the
literature must be clarified.

The last development stage (i.e. after 7~Gyr) of \ABAB\ is
morphologically close to SB0 galaxies, and so it is morphologically
comparable to the \citet{2004ApJ...603..495P} observations. We can thus
determine the amount of gas in a region of similar surface limited by
the beam size. For ten objects, the radius encircling the mass that
has been measured ranges from 0.3 kpc to 2.6 kpc, but for their SB0
subsample (NGC\,2859, NGC\,2950, NGC\,3081, NGC\,4340, and NGC\,4371)
$r < 1$~kpc and $r < 2.3$~kpc are more representative. For $r <
1$~kpc, observational gas mass upper limits range from 1.5\,10$^7$ to
8.1\,10$^7$ \msol\ , whereas it amounts to 4.4\,10$^7$~\msol\ for
\ABAB. For $r < 2.3$~kpc it increases to 2.8\,10$^8$~\msol\ for our
simulation, whereas for NGC\,3081 the gas mass amounts to 6.2 or
6.5\,10$^8$~\msol. Since our simulation is not fine tuned to fit any
particular galaxy, we can conclude that \ABAB\ is well within the
observational constraints. A value of 4.4\,10$^7$~\msol\ is  low enough to
explain why \citet{2004ApJ...603..495P} were unable to detect
molecular gas in NGC\,2859, NGC\,4340, and NGC\,4371, but obviously one
cannot conclude that the gas component do not play a role in the
formation of nuclear bars.

\subsection{Stellar population in nuclear bars}
\label{ssec:stellarpop}

Studying the stellar populations of double-barred galaxies NGC\,2859,
NGC\,3941, NGC\,4725, and NGC\,5850, \citet{2013MNRAS.431.2397D} have
recently concluded that the inner bars are younger and more metal rich
than the large-scale primary bars. It has been argued
\citep{2007A&A...465L...1W} that the stellar age distribution in bars
has to be interpreted with a dynamical approach. For instance, two
regions of apparent low age at the end of the large-scale bars
\citep{2007A&A...465L...9P} are due to the accumulation of a composite
stellar population, younger in average and 
trapped in orbits shaped
like an ellipse in average, aligned with the bar.

To make a comparison with \citet{2013MNRAS.431.2397D}, a mean age
profile has been computed as the average of individual particle age in
annuli 100~pc wide and $\pm 100$~pc thick centred on the
nucleus. Being an azimuthal average approach, each annulus mixes the
population inside and outside the structures (inner and outer bars,
circumnuclear ring).  Moreover, all particles of the initial
population  arbitrarily have the same age (0 at the beginning of the
simulation) so that snapshot time is also the age of the initial
population.

\begin{figure}
\centering
\includegraphics[keepaspectratio,width=\hsize]{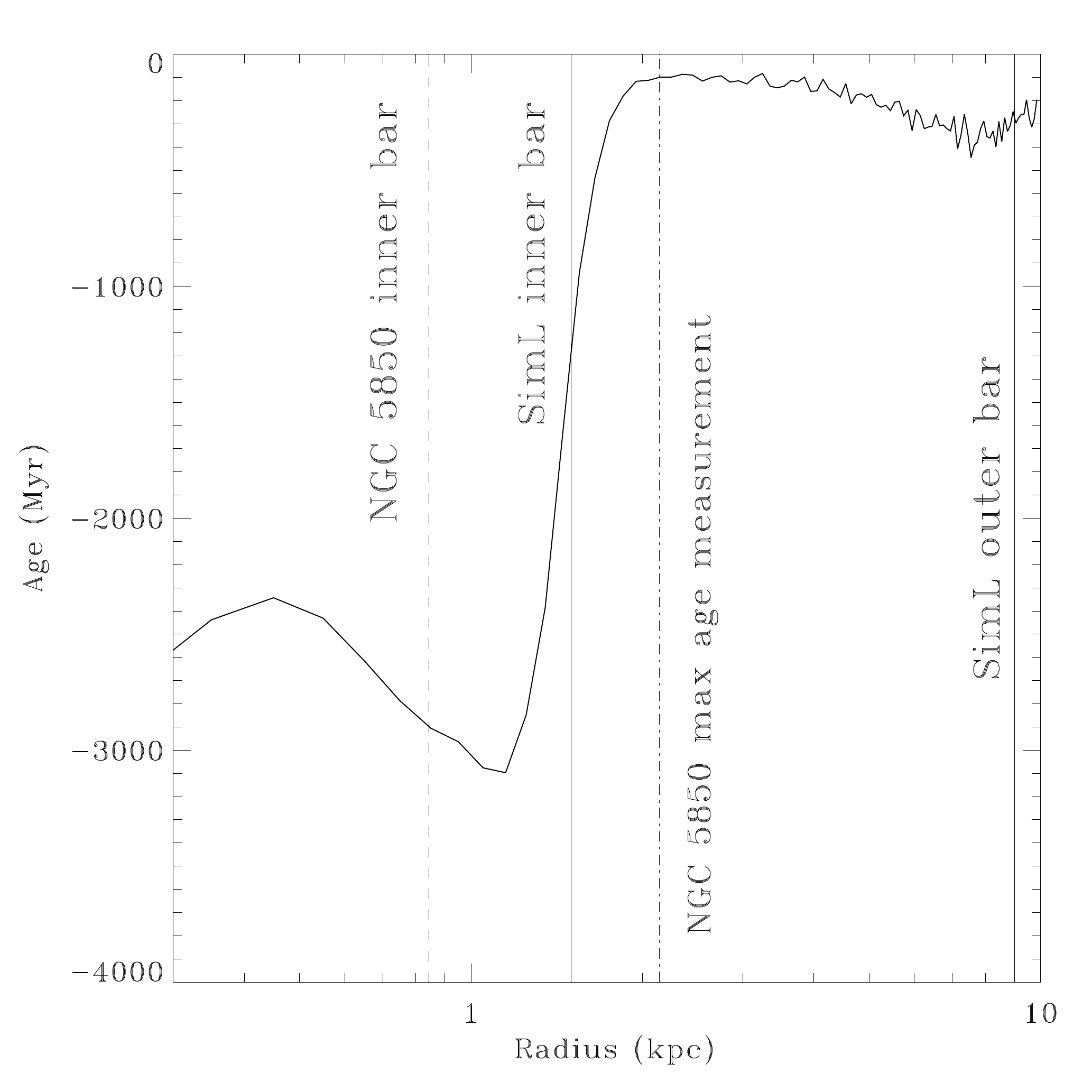}
\caption{Radial profile of the age difference with the time of the
  snapshot ($t = 9.486$~Gyr) for \ABAB. The mean age has been computed
  in concentric annuli of 100~pc wide. For the sake of comparison, the
  radius of the  NGC\,5850 inner bar and the limit of age measurement
  \citep{2013MNRAS.431.2397D} have been scaled to \ABAB\ and
  overplotted as  dashed and dot-dashed lines, respectively.}
\label{fig:periodage}
\end{figure}

In Fig.~\ref{fig:periodage} we show the relative mean age of the
particle population at $t = 9.486$~Gyr.  The region inside the nuclear
bar is 2.5$-$3.0~Gyr younger than the large-scale bar. Because of the
mixing with older regions outside the inner bar, and the rough
assumption that all initial particles have an age of  9.486~Gyr, this value is
certainly a lower limit. The outermost region of the large-scale bar
is also younger. This effect is obviouly that discovered by
\citet{2007A&A...465L...9P} and explained by
\citet{2007A&A...465L...1W}.

Even if a direct and quantitative comparison with
\citet{2013MNRAS.431.2397D} is not possible because the averaging
technique is different (they measured ages along ellipses fitted on
the isochrone maps), the results are in good qualitative agreement. In
particular, the age difference for NGC\,5850 amounts to $\approx
4$~Gyr between the nuclear bar and the outermost measurements.  To
help the comparison, the large-scale bar radius of NGC\,5850 (using
the   \citet{2013MNRAS.431.2397D} value, i.e. 63\arcsec) has been
scaled to \ABAB\ ($l_\mathrm{p} \approx 9$~kpc). The large-scale bar
radius was used as its determination is much more accurate than
the nuclear radius. In Fig.~\ref{fig:periodage} the outermost limit of
age measurement by \citet{2013MNRAS.431.2397D} is also plotted showing
that their age measurements might be well inside the large-scale bar.

\subsection{Pattern speeds}
\label{ssec:pattern}

\begin{figure}
\centering
\includegraphics[keepaspectratio,width=\hsize]{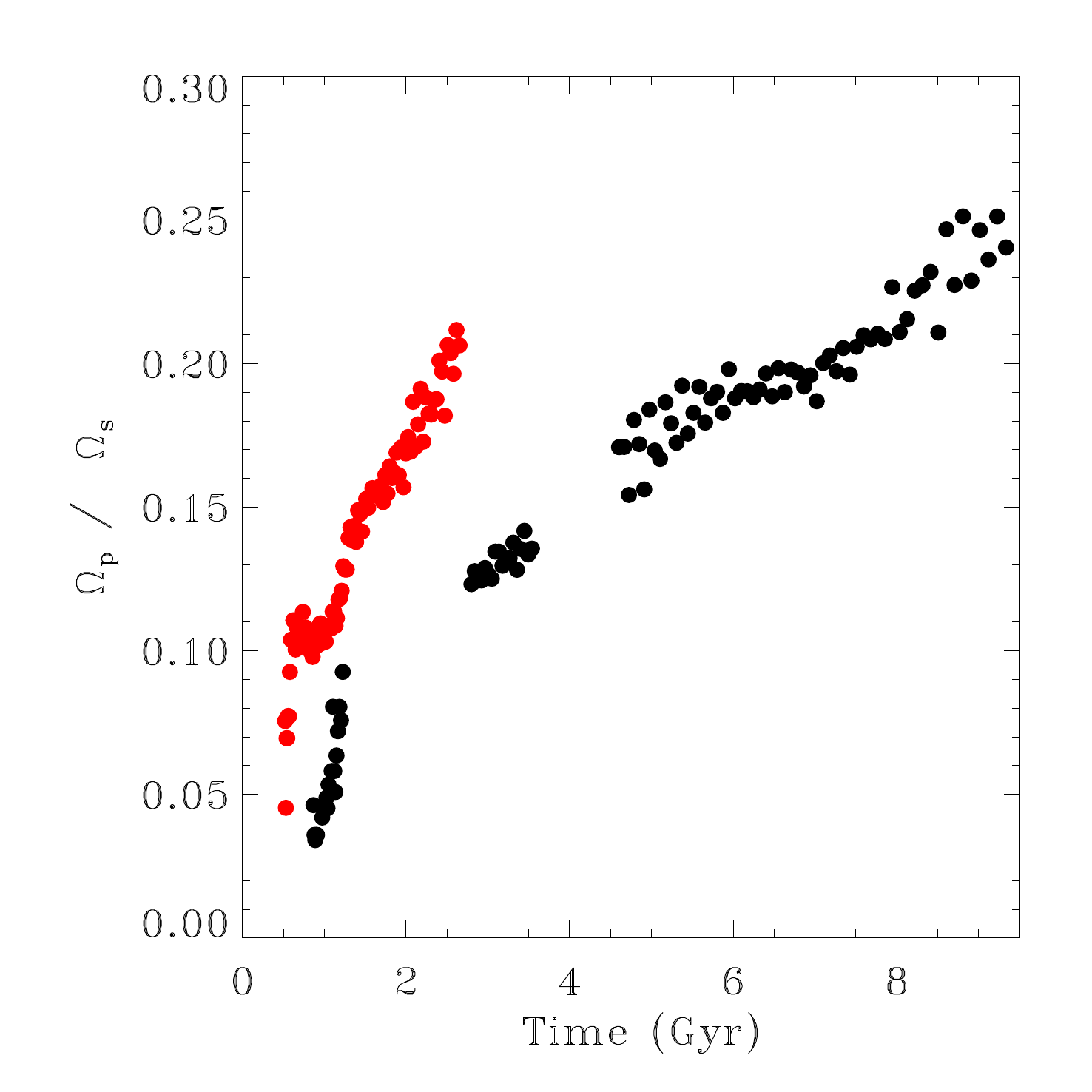}
\caption{Ratio of the primary to secondary bar pattern speed. Black
  dots are for \ABAB, red dots for \FBFC.}
\label{fig:period_ratio}
\end{figure}

The ratio of pattern speeds (\Omegap/\Omegas\ or equivalently \Ts/\Tp,
cf. Fig~\ref{fig:period_ratio}) is not constant over time. For the two
simulations, both \Omegap\ and \Omegas\ decrease as a function of
time (Figs.~\ref{fig:FBFC_period} and \ref{fig:ABAB_period}) but are
not locked into a particular ratio.  The slopes seem different between
\FBFC\ and \ABAB, but are in fact consistent with a factor of
roughly two in dynamical timescales. This means that the ratio is the
same for \ABAB\ after twice the time needed by \FBFC. Because on average
\Omegas\ decreases much faster than \Omegap, the ratio
\Omegap/\Omegas\ monotonically increases up to $\approx
0.21-0.25$. Only a change in the slope and a lower dispersion between
6 and 8~Gyr are notable for \ABAB, mainly due to \Omegas\ variations
during the circumnuclear ring collapse (cf. Sect~\ref{ssec:abab}).
The fluctuations of the ratio is due to oscillations of \Omegas\ and
\Omegap\ (Figs.~\ref{fig:FBFC_period} and \ref{fig:ABAB_period}) that
depend on the relative phase of the two bars, a fact already observed
by \citet{2004ApJ...617L.115E} and \cite{2007ApJ...654L.127D}.

The best model of \citet{1993A&A...277...27F} has $\Omega_p/\Omega_s
\approx 0.32$, and $0.3$ for \citet{2004ApJ...617L.115E}, values that
our simulations never reach. Our values of \Omegap\ and \Omegas\ are
 comparable, however, to the few measurements in double-barred galaxies.
Most observational methods to directly estimate $\Omega_p$ cannot be
easily applied to $\Omega_s$ and, more generally, are not suited when
multiple patterns are present
\citep{2006MNRAS.371..451M,2009ApJ...690..758S}.  However, keeping in mind the
questionable viability of $\Omega_s$ measurement, we must  rely
on available attempts.

Therefore, if we assume two bars in NGC\,5248 \citep[cf. discussion
  in][]{2013A&A...556A..98V},
\Omegap$\,=30$~\omegaunits\ \citep{2002ApJ...575..156J} and
\Omegas$\ga 135$~\omegaunits\ leading to a ratio \Omegap/\Omegas~$\la
0.23$.  \citet{2003ApJ...599L..29C} found a value as low as 0.13 for
NGC\,2950 using the Tremaine-Weinberg method \citep[but
  see][]{2006MNRAS.371..451M,2009ApJ...690..758S}.
\citet{2005ApJ...632..253H} and \citet{2009ApJ...704.1657F} applied
the same technique to H$\alpha$ velocity fields to determine pattern
speeds and found $0.3 < \Omega_p/\Omega_s < 0.55,$ but with large error
bars on \Omegas. An in-depth analysis of NGC\,1068, based both on
multiwavelength observations and numerical simulations, led
\citet{2006MNRAS.365..367E} to conclude that $0.19 \la
\Omega_p/\Omega_s \la 0.26$.

\citet{2007AJ....133.2584Z} introduced a potential-density phase-shift
method that is able to determine the position of multiple corotations
in a galaxy without any kinematical information. This method is thus
particularly well suited to double-barred galaxies. Moreover, it is
based on the properties of quasi-stationary density wave
modes. \citet{2007AJ....133.2584Z} did not publish values for
$\Omega_p$ and $\Omega_s$, but only radii for \CRs\ and \CRp. However,
the ratio of corotation radii \CRs/\CRp\ can be used as it evolves
like $\Omega_p/\Omega_s$. For their small sample (seven objects), they
found ratios between 0.086 (NGC\,1530) and 0.3 (NGC\,936) and a mean
of $\approx 0.17$. For \FBFC\ and \ABAB, \CRs/\CRp\ starts at $\approx
0.03$ when a nuclear bar is just detected for the first time and then
increases  to respectively $\approx 0.24$ and $0.30$. Our values are
thus comparable even for SB0 galaxies.

The most recent survey \citep{2014MNRAS.444L..85F} of pattern speeds
in double-barred galaxies, using a phase reversal technique in
H$\alpha$ kinematical maps \citep{2014ApJS..210....2F} found a narrow
range of ratios $0.29^{+0.05}_{-0.05}$, a bit higher than our end
values.  However, \citet{2014MNRAS.444L..85F} have introduced an
important assumption: the nuclear bar ends near its corotation (\CRs)
which in turn overlaps (or is close to) the primary bar ILR
(\ILRp). If this new technique could be useful to determine
the location of resonances (although a firm physical foundation must be found), it gives no information on the type of these resonances, especially the main ones,  ILR, ultraharmonic (UHR, also called 4:1), CR, and OLR. Therefore, the
identification of the resonance needs to assume that any bar ends
inside and close to its corotation. Whereas this is reasonable true
for the primary bars (although it has been argued by
\citet{2006A&A...452...97M} that typical observational bar length
measurement techniques are more correlated with the  UHR resonance than with the CR), this assumption is questionable for
nuclear bars. Indeed, if \CRs\ is determined in \ABAB\ as being the
end of the nuclear bar, and moreover identical to \ILRp, this leads to
a large underestimation of \Omegas. For the sake of illustration, 
we consider the moment $t=7.5$~Gyr. At this time, $R_\mathrm{ILRp}
\approx 4.84$~kpc and $\Omega_p\approx 15$~\omegaunits. If we assume
that this radius is identical to the corotation radius of the nuclear
bar, this leads to $\Omega_s\approx 45$~\omegaunits\ instead of
$\approx 73$~\omegaunits\ (see also Fig.~\ref{fig:fft} in the next
section). Therefore, the ratio \Omegap/\Omegas\ might be spuriously
estimated to be 0.33, a value which appears to be in good agreement
with \citet{2014MNRAS.444L..85F}, instead of $\approx 0.2$, its
\emph{\emph{real}} value.

\subsection{Bar lengths}
\label{ssec:shape}

The absolute sizes of simulated secondary bars fit quite well with the
values given by \citet{2011MSAIS..18..145E}. Erwin showed that
$l_\mathrm{s}/l_\mathrm{p}$, the secondary over primary bar lengths,
ranges from 0.025 to 0.25 with a median of $\approx 0.12$. For the
first nuclear bar of \ABAB, $l_\mathrm{s}/l_\mathrm{p}$ remains
roughly constant at 0.05. It is only when the nuclear bar has been
fully dissolved that this ratio reaches 0.11 (this is then the size of
the remnant nuclear disc relative to the large-scale bar). The inner
bar restarts with a rather high ratio (0.275 at $t\approx 4.5$~Gyr),
but then decreases to 0.15-0.19. However, the observational relative
size distribution does not show any strong evidence of two different
kinds of secondary bars that could help to distinguish the first
generation from the subsequent ones. Our simulations should also
show  some similarities with that of \citet{1993A&A...277...27F} since our
recipes are quite similar, although our initial conditions are
different. Indeed, the basic process leading to the formation of a
nuclear bar is the same. However, their best model has a bar length
ratio of $0.26$.

Although it is incorrect to base an argument on absolute sizes, 
it is also
noteworthy that, all other factors being equal, if primary bars have
low \Omegap\ (as for \ABAB), they are  longer than bars
with high \Omegap\ (as for \FBFC), assuming here that they
systematically fill the region inside their corotation radius. 
\citet{2011MSAIS..18..145E} emphasizes that the sizes of large-scale
bars are longer in double-barred galaxies than in single-barred ones
whereas all the other properties are  similar.

\section{Discussion}
\label{sec:discussion}

\FBFC\ and \ABAB\ having morphological, kinematical and stellar
population properties similar to real galaxies, we can now discuss
their dynamical structure and long-term evolution.

\subsection{Resonance overlap or not?}
\label{ssec:mode}

\begin{figure}
\centering
\includegraphics[keepaspectratio,width=\hsize]{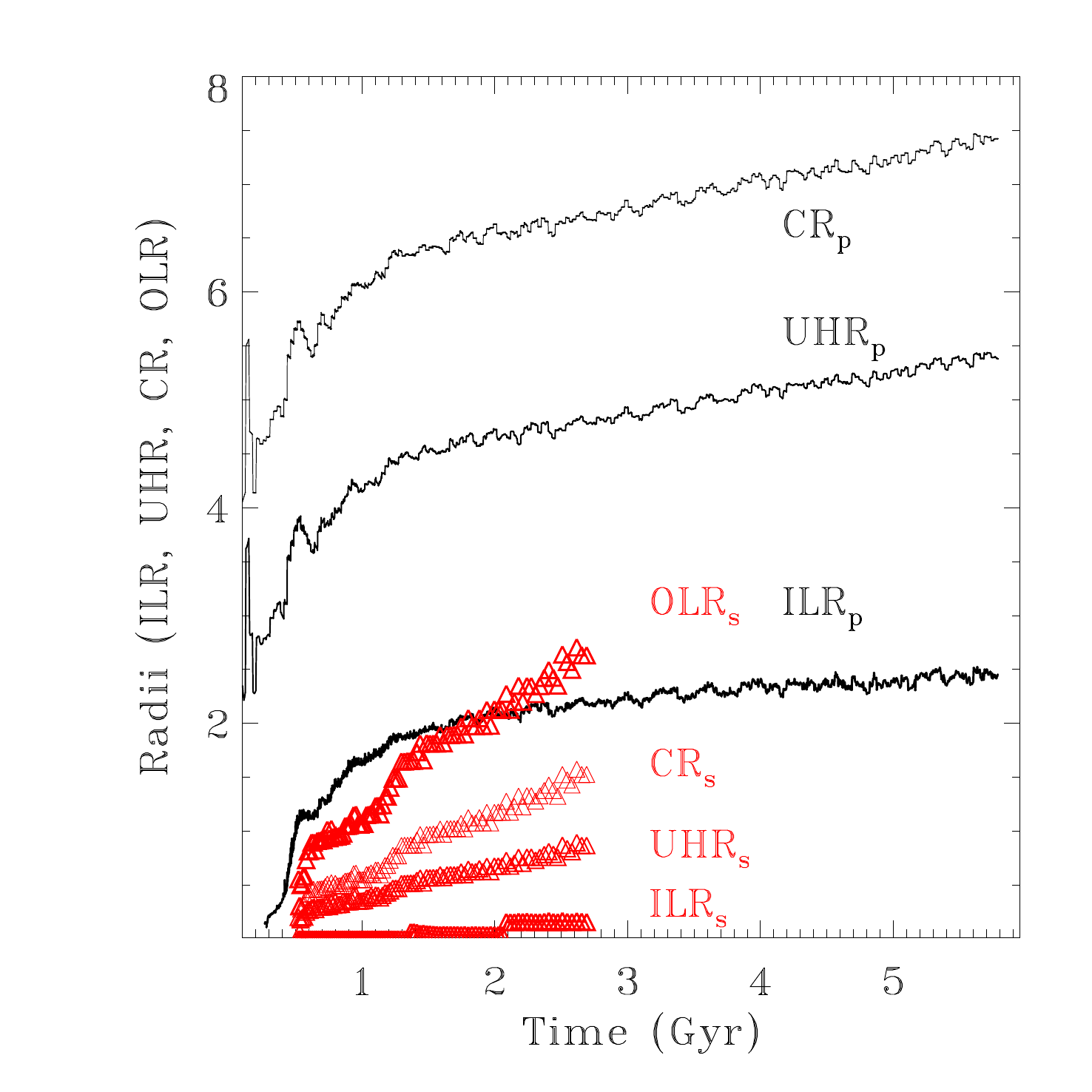}
\includegraphics[keepaspectratio,width=\hsize]{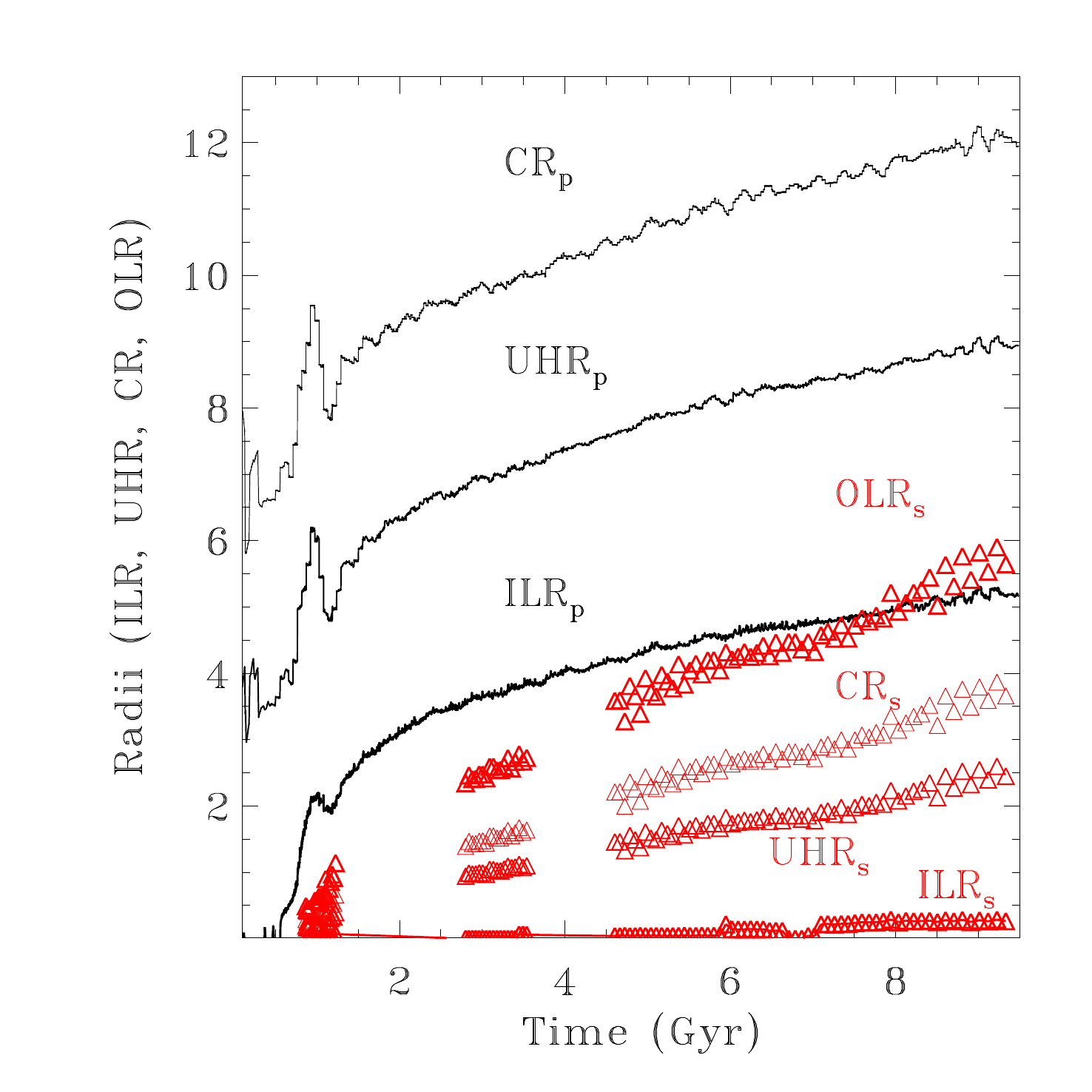}
\caption{Radius of the linear resonances (ILR, UHR, CR, OLR) for the
  secondary bar (subscript 's', red triangles) and the primary bar ('p', black
  dots) in the simulations \FBFC\ (top) and \ABAB\ (bottom). \OLRp\ is
  not shown. }
\label{fig:pos_resonance}
\end{figure}

In the framework of the
epicyclic approximation, we  solved the equations $\Omega-\kappa/n
= \Omega_{p,s}$ for the resonance radii (\ILRp\ and \ILRs\ for
$n=2$, \UHRp\ and \UHRs\ for $n=4$, \CRp\ and \CRs\ for $n=\infty$,
\OLRp\ and \OLRs\ for $n=-2$). \citet{1992MNRAS.259..328A} and
\citet{2006A&A...452...97M} have shown that, even in the case of
strong bars for which the epicyclic approximation breaks down, the
errors on the resonance positions remain within 10\%, especially in
the case of the ILR and the CR. In Fig.~\ref{fig:pos_resonance} we
show the evolution of ILR, UHR, CR, and OLR as a function of
time. Since most of the time two ILRs exist, hereafter we only deal  with the
outermost one (named oILR by most authors).  Moreover, because it is very
close to the centre in the case of \FBFC, the position of \ILRs\ is
not always reliable for $t\la 2$~Gyr.

The immediate observation is that the whole resonance system of the
inner bar, up to \OLRs, remains well inside the \ILRp\ for a long time
(roughly 2~Gyr for \FBFC\ and 8~Gyr for \ABAB). Therefore, there is no
\CRs--\ILRp\ overlap, unlike \citet{1993A&A...277...27F} and many
other  similar studies that followed. 

However, such a situation is not new
as a few examples have been already reported by
\citet{1999A&A...348..737R} and \citet{2000A&A...362..465R}, but with
different experimental setups (either 2D simulations or an analytical
bulge component).  Moreover, \citet{1995A&A...301..359M} reported
evidence of an \OLRs--\ILRp\ coupling in the case of NGC\,4736, before
the paradigm of \CRs--\ILRp\ coupling was firmly established. They
claimed this configuration could weaken the interaction and increase
the lifetime of the nuclear bar.

To further investigate the dynamical impact of possible resonance
overlaps, we have built spectrograms, as introduced by
\citet{1986MNRAS.221..195S} and widely used \citep[see e.g.][and
  references therein]{2011MNRAS.417..762Q}. We note $m$ the
azimuthal wavenumber of a wave of frequency $\omega$. Spectrograms
can be used to determine the pattern speed $\Omega$ of any wave as being
$\Omega = \omega/m$ and helps to identify any potential non-linear
mode coupling when resonances overlap.  As discussed by
\citet{1999A&A...348..737R} and \citet{2011MNRAS.417..762Q},
spectrograms show many small features, often transient, probably
linked to the main modes. Some of these features with a well-defined
pattern speed can be beat modes triggered by the non-linear couplings
of other waves at the location of resonance
overlaps. \citet{1999A&A...348..737R} found many kinds of resonance
overlaps in addition to the classical \CRs--\ILRp, although they were not
always able to determine whether these couplings were stationary or
accidental.

\begin{figure}
\centering
\includegraphics[keepaspectratio,width=\hsize]{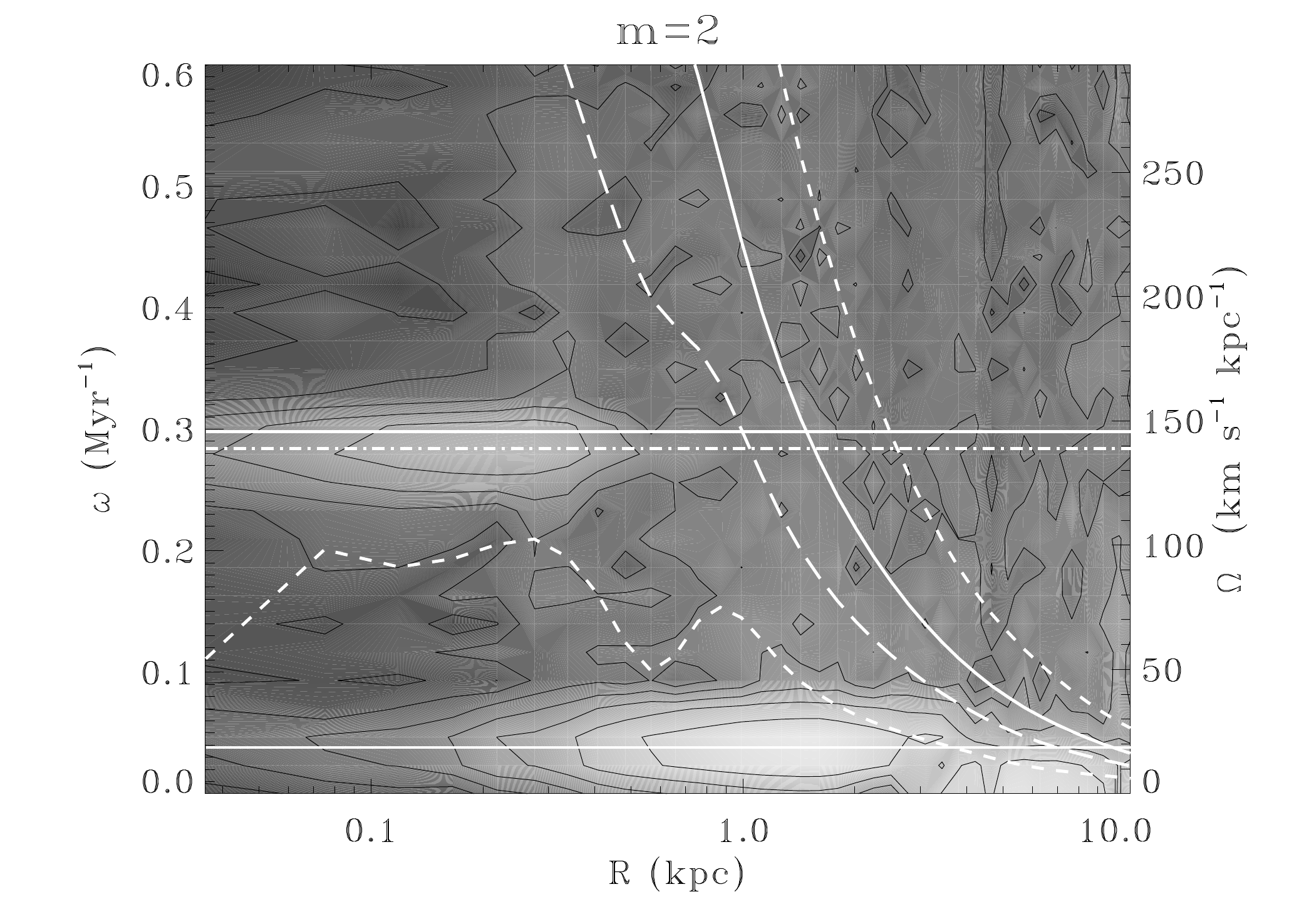}
\includegraphics[keepaspectratio,width=\hsize]{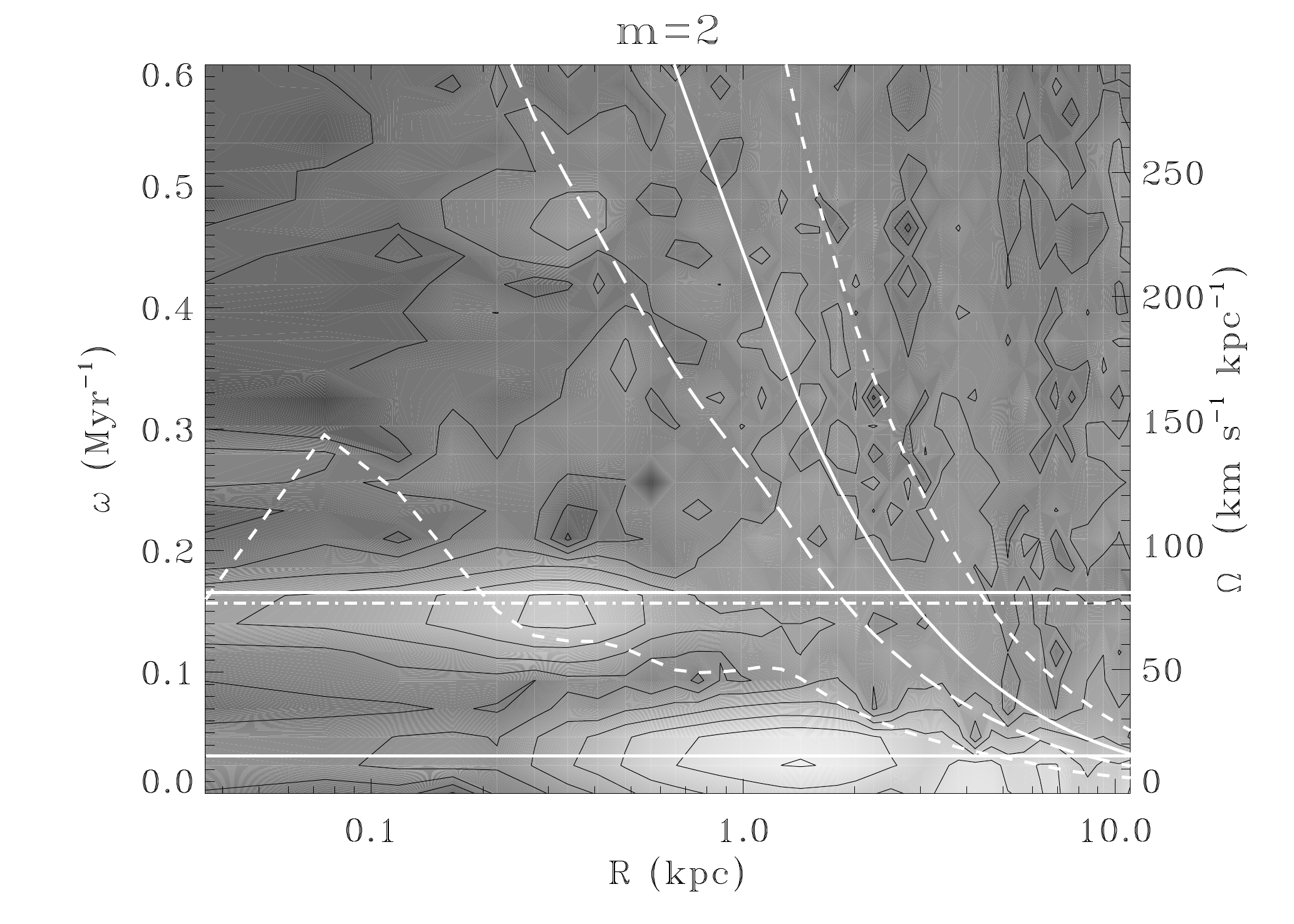}
\caption{$m=2$ power spectra for \ABAB\ in $\log$ scale. The time
  window spans 269~Myr centred at $t=3.3$ and $t=7.5$~Gyr. The left
  vertical scale gives values of $\omega$, the wave frequency in
  Myr$^{-1}$, whereas the right vertical scale is for the pattern
  speed $\Omega=\omega/2$ in \omegaunits. The horizontal white lines
  represent \Omegas\ and \Omegap, respectively, at the beginning of the
  time window (full line) and the end (dot-dashed
  one). \Omegas\ decreases from 146 to 139 at $t=3.3$~Gyr and from 81
  to 71~\omegaunits\ at $t=7.5$~Gyr, whereas \Omegap\ decreases by
  less than 1~\omegaunits\ so that the two lines cannot be
  distinguished. The averaged curves $2\Omega -\kappa$ (which allows
  the ILR to be identified) and $2\Omega +\kappa$ (for the OLR) are drawn
  as white short dashed lines, $2\Omega -\kappa/2$ (for the UHR) as
  long dashed line, and $2\Omega$ as a solid line (for the CR). A
  logarithmic scale has been chosen for the radius to emphasize
  the central kpc. }
\label{fig:fft}
\end{figure}

Figure~\ref{fig:fft} shows spectrograms for $m=2$, centred at $t=3$ and
$7.5$~Gyr for \ABAB\ and computed in a window of 269~Myr corresponding
to 256 outputs. To compute the spectra we  included only the
stellar component to extract the information on stellar modes. Similar
spectrograms have been obtained for \FBFC. Spectra with greater
wavenumbers ($m \ge 4$) confirm what can be inferred from $m=2$, in
particular the presence of harmonics of the large-scale bar mode. They
are not shown here. For the two moments shown in
Fig.~\ref{fig:fft} representative of the rest of the simulations, the
two dominant modes correspond to the secondary and primary bars, with
pattern speed values in good agreement with all other measurements
(cf. Sect.~\ref{sec:model}). The features are not as thin as for
\citet{1997A&A...322..442M}. This is due in part to our narrower time
window. This also shows that these features do not correspond to
quasi-stationary structures.

Figure~\ref{fig:fft} clearly shows that most of the power of the nuclear
bar mode remains inside the \UHRs. For the large-scale primary bar, it
has been argued that the most appropriate orbits to sustain a bar are
those existing inside the UHR \citep[see e.g. Sect.~6.1~of][
  and reference therein]{2006A&A...452...97M}. This also seems  to be
true for the secondary bars of \ABAB\ and \FBFC. This additionally
supports our previous claim (Sect.~\ref{ssec:pattern}) that one cannot
systematically assume that the morphological end of a nuclear bar
coincides with \CRs.

Finally, whatever the approach, the linear resonance analysis or the
mode analysis both lead to the same conclusion: there is no resonance
overlap or mode coupling in \ABAB\ and \FBFC.

\subsection{Double-bar lifetime}
\label{ssec:lifetime}

\begin{figure}
\centering
\includegraphics[keepaspectratio,width=\hsize]{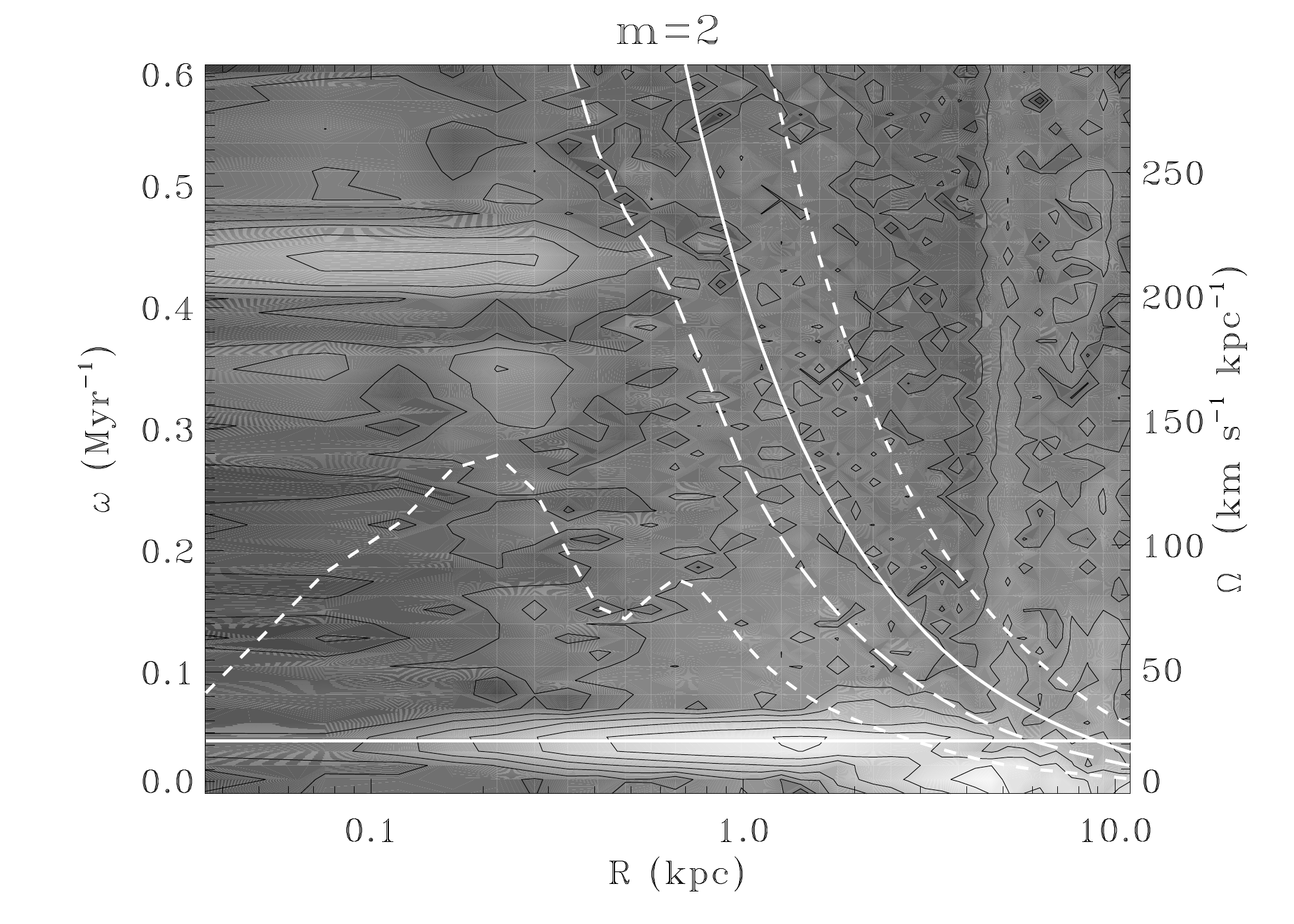}
\includegraphics[keepaspectratio,width=\hsize]{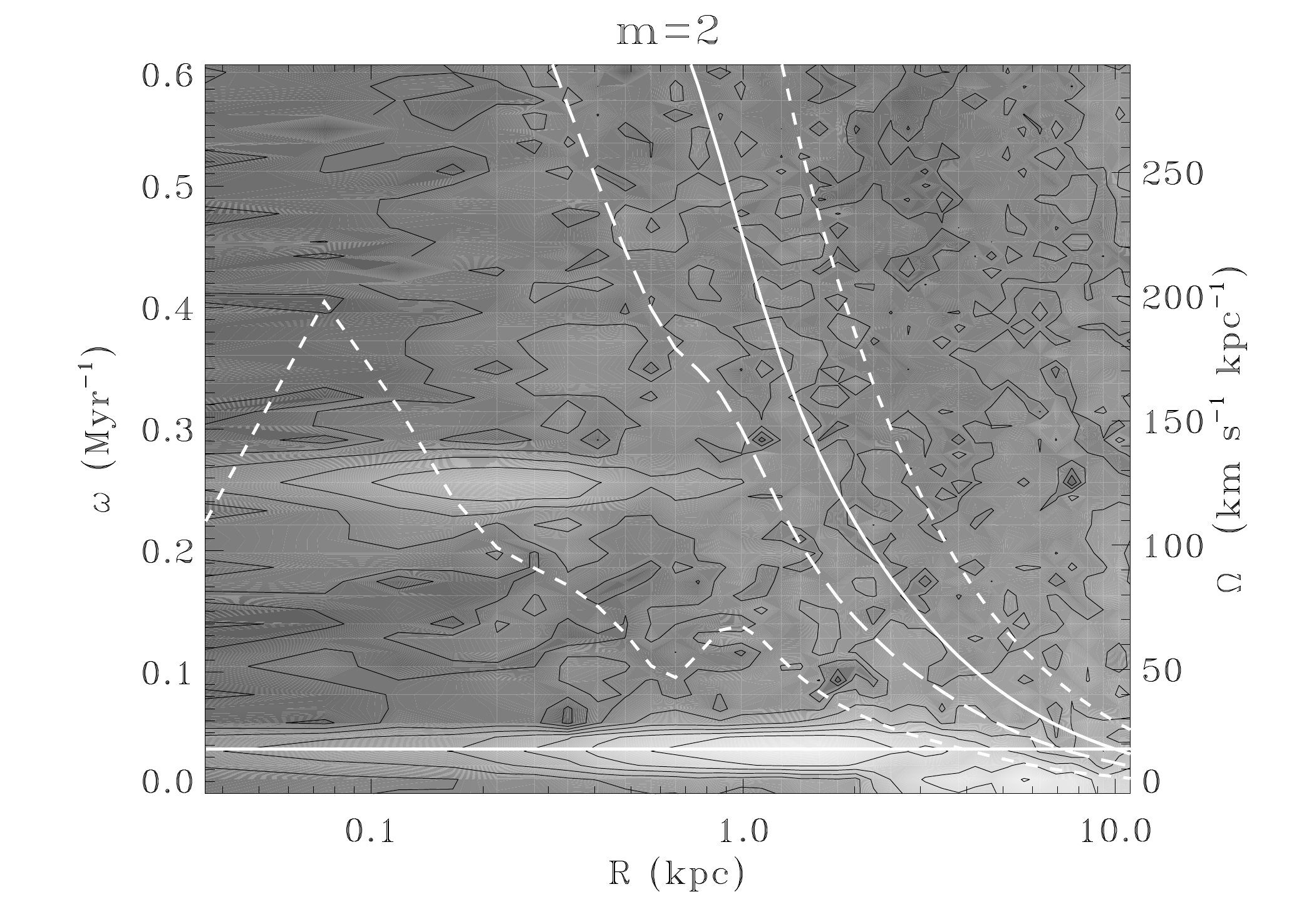}
\caption{$m=2$ power spectra for \ABAB\ centred at $t=2$ (top panel)
  and $t=4.275$~Gyr (bottom) in a window of 538~Myr, whereas the inner
  bar cannot be detected on density maps. Consequently,
  \Omegas\ cannot be measured because the position-angle of the central
  structure cannot be properly determined. }
\label{fig:fft2}
\end{figure}

\citet{2002MNRAS.337.1233R} reported the case of a long-lived double-barred
system (their {\tt model I}) which could be very similar to
\ABAB, although  \ABAB\ is 3D and fully self-consistent whereas their
{\tt model I} is 2D and the bulge is represented by an analytical Plummer
sphere. However, the fact that long-lived systems can exist is 
important in order to explain the high frequency of double-barred galaxies. In
our case, the nuclear bar lifetime is clearly linked to the lifetime
of the central bar mode. The fact that we are not able to
determine a proper pattern speed during roughly 1~Gyr centred on
$t\approx 2$ and 4~Gyr (cf. Fig.~\ref{fig:abab}) does not mean that
\emph{dynamically} the nuclear mode has disappeared. Indeed, the mode
still exists (Fig.~\ref{fig:fft2}) at a pattern speed of respectively
$\approx 215$ and 125~\omegaunits. This means that even if, on
photometrical grounds, it can be stated that the nuclear bar has been
dissolved into a triaxial bulge or pseudobulge
\citep{2004ARA&A..42..603K}, or even a weaker structure like nuclear
spirals (cf.~Sect.~\ref{sec:evol}), the bar mode does survive, which  in
turn facilitates the morphological revival of the nuclear bar later on.

\citet{2007ApJ...654L.127D} also reported long-lived double-barred
systems in purely collisionless simulations. A direct comparison is
 not possible since the initial conditions of their
simulations, especially the bulge part, were fine-tuned to produce
such gas-free double-barred galaxies. However, their enforced rotating
bulge mimics perfectly the effect of a dissolved or weak first nuclear
bar.

Another crucial dynamical feature of our models that could play an
important role in the lifetime of the secondary bar, is the absence of
any overlap of resonances for a long time.  Indeed, the whole nuclear
dynamical system evolves as if  embedded in a quasi-stationary
gravitational background due to the rest of the galaxy. The mass (both
gas and stars) encircled by the \OLRs\ is so large that the two
dynamical systems are unable to couple together through their
resonances. For \ABAB, almost 38~\%\ of the total mass of the galaxy
lie inside $R \la 5$~kpc at $t=7$~Gyr.

All previous works that have looked for mode coupling in 3D
self-consistent N$-$body/hydrodynamical simulations have found an
\CRs--\ILRp\ overlap whereas their double-barred systems were
short-lived \citep[e.g.][]{1993A&A...277...27F}. One noticeable
exception is \citet{2002MNRAS.337.1233R} who argue that this mode
coupling is not necessary for the coexistence of bars and show
examples of other kinds of coupling. \citet{2000A&A...362..465R},
using 2D N$-$body simulations and inelastically colliding massless gas
particles, exhibit an example very close to \ABAB\ with \OLRs\ well
inside \ILRp\ (their {\tt model A2.5}). This is the situation of \FBFC\ for $t\la
2$~Gyr and \ABAB\ for $t\la 8$~Gyr.  However, their model develops a
very weak primary bar (more oval than \ABAB) and the $m=2$ of the
nuclear mode remains close to the maximum of $\Omega-\kappa/2$ all
through the simulation, which is not the case of our simulations.

\begin{figure}
\centering
\includegraphics[keepaspectratio,width=\hsize]{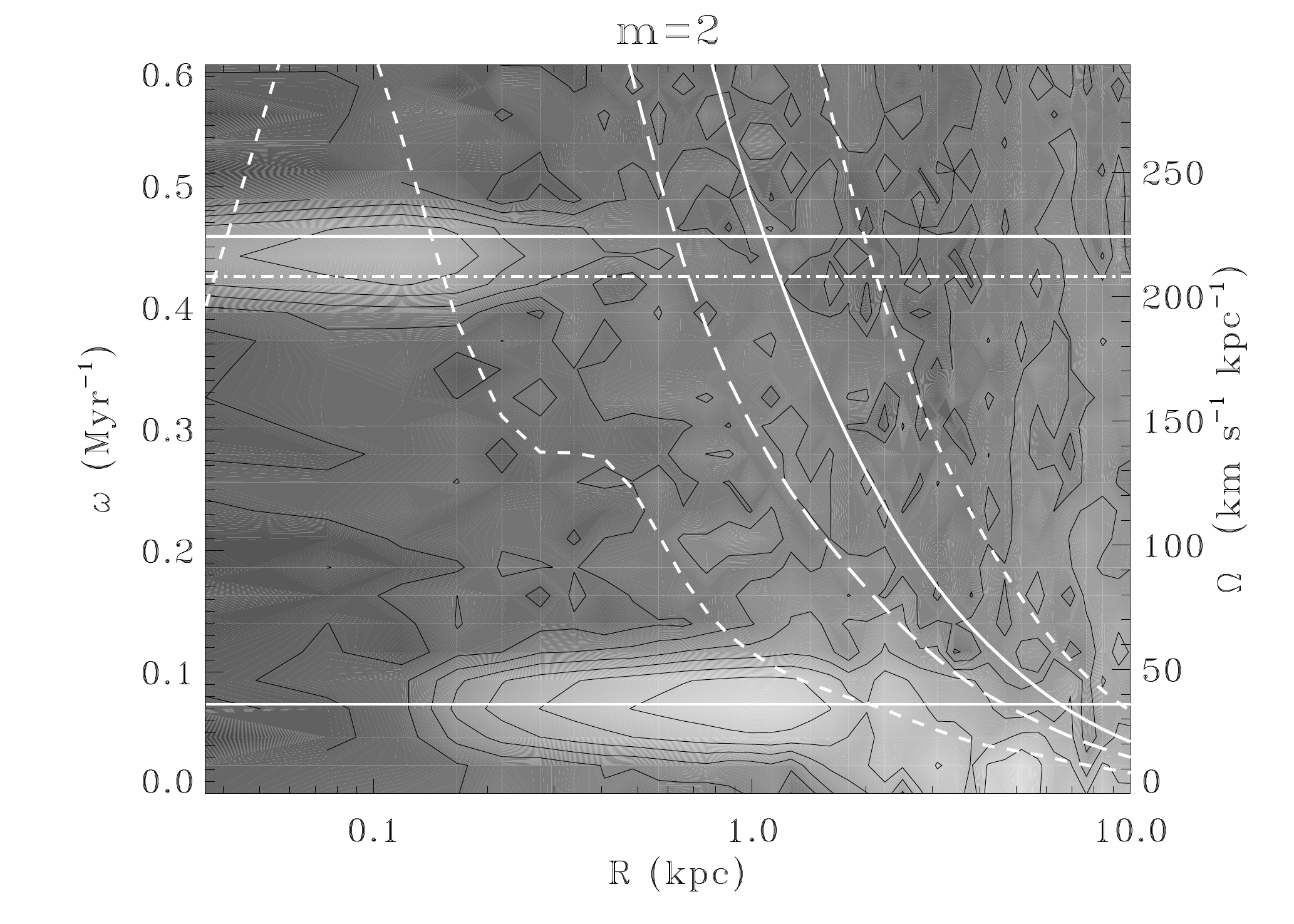}
\includegraphics[keepaspectratio,width=\hsize]{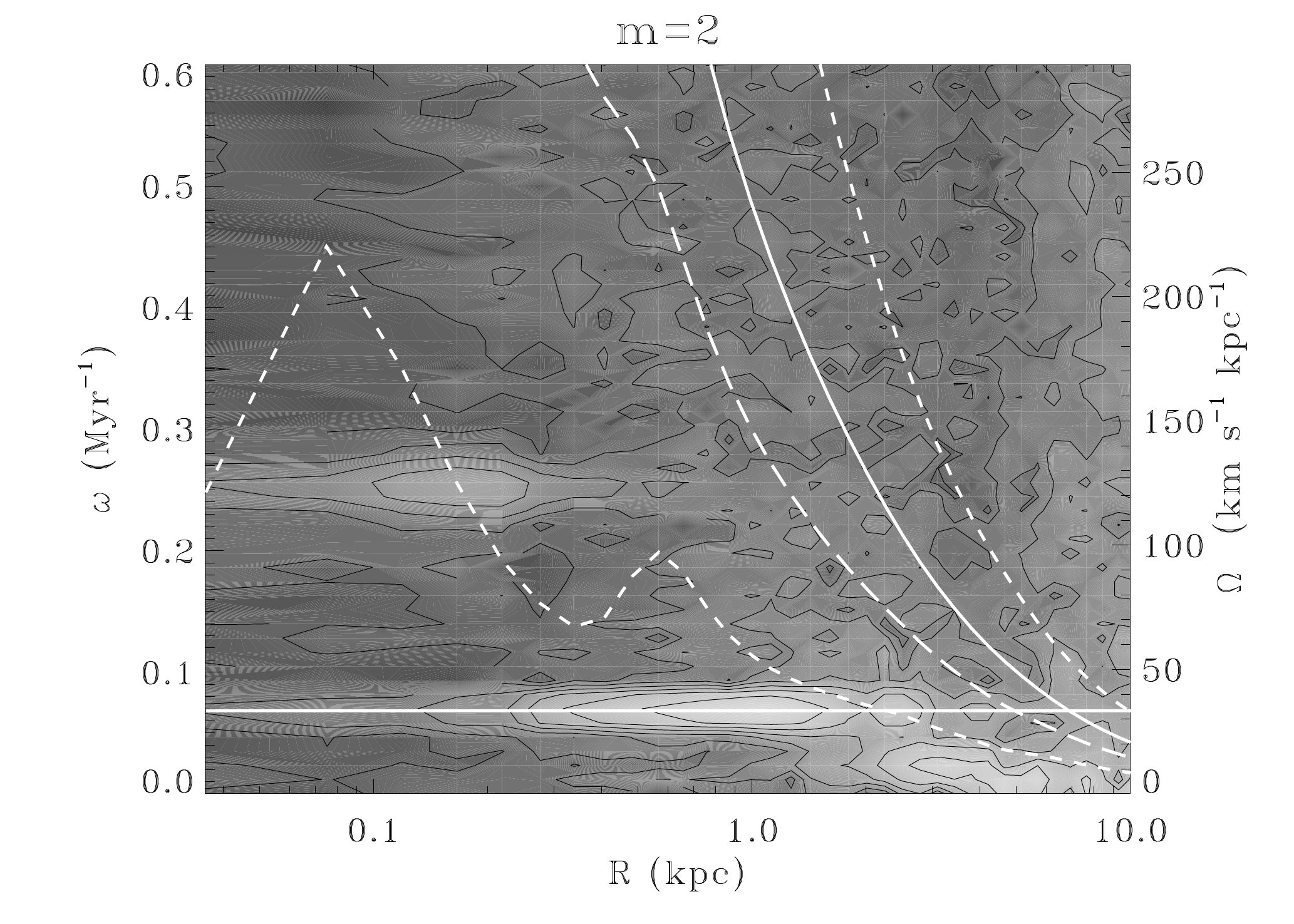}
\includegraphics[keepaspectratio,width=\hsize]{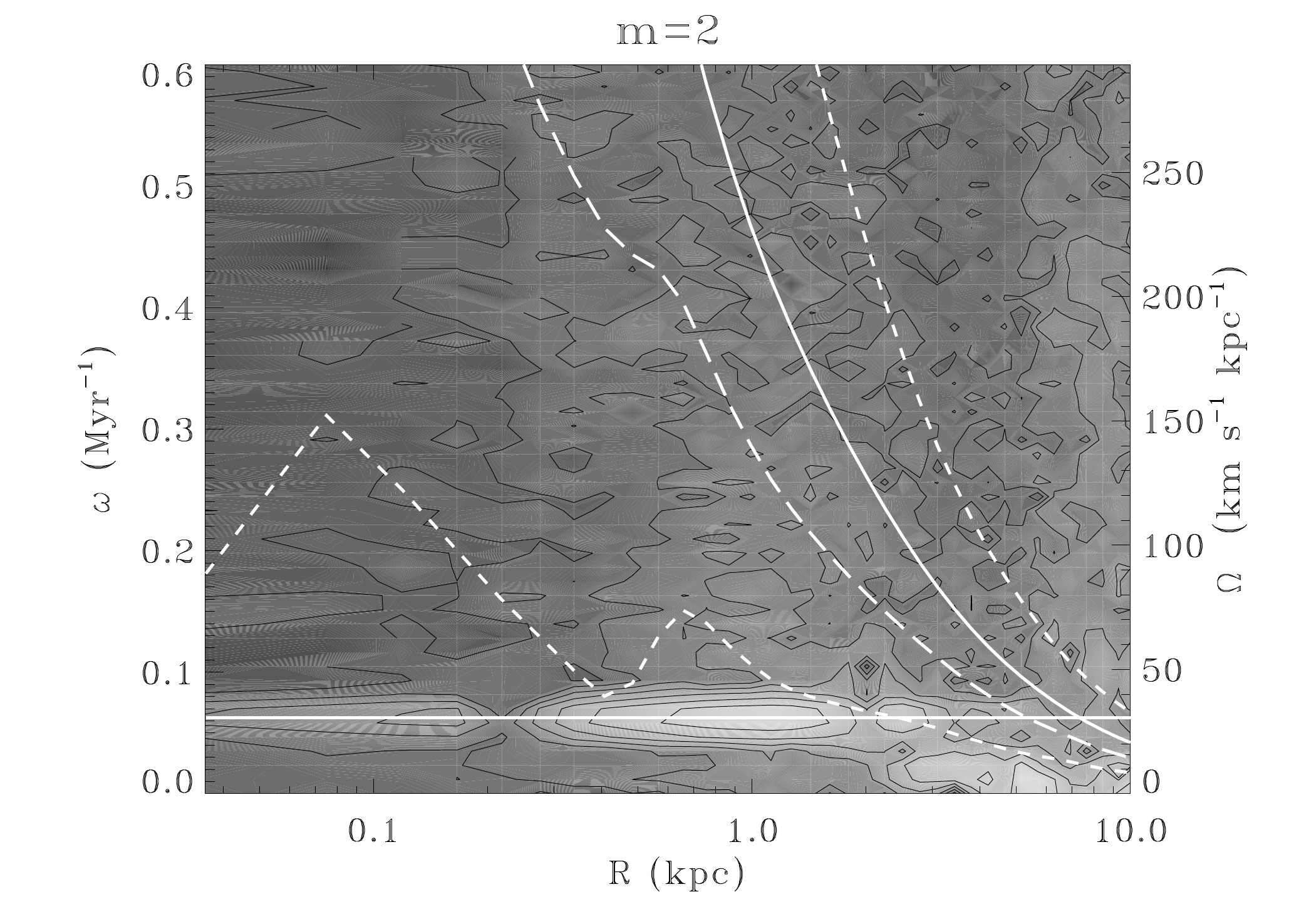}
\caption{$m=2$ power spectra for \FBFC\ centred at $t=2$ (top panel), when
  \OLRs\ intersects \ILRp, $t=3.4$ (middle), and $t=5.47$~Gyr (bottom)
  where the nuclear bar is dissolved. For $t=2$~Gyr the time window
  has been halved (269~Myr) because \Omegas\ evolves rapidly.}
\label{fig:fft3}
\end{figure}

Can we infer a strong dynamical perturbation when \OLRs\ position
comes close to \ILRp\ and eventually overlaps with \ILRp? For \FBFC, less
than 1~Gyr after \OLRs\ intersects \ILRp, the nuclear bar is
transformed into a disc with a very time-dependent shape. It is
also remarkable that the nuclear bar then starts to dissolve in
the case of \FBFC\ whereas it continues to evolve as a large oval for
\ABAB, showing larger fluctuations in \Omegas. Three spectrograms centred at
$t=2, 3.4,$ and $5.47$~Gyr for \FBFC\ (Fig.~\ref{fig:fft3}) show that the nuclear
bar mode progressively vanishes whereas the large-scale bar mode (with
a lower pattern speed) expands toward the centre. However, the
\OLRs\--\ILRp\ overlap seems to play no role. Indeed, the primary
bar mode is at a maximum just inside \ILRs\ , but we have not found a mode
associated with the nuclear bar that might be present near the
\OLRs\ to explain the energy and angular momentum exchanges. The
overlap thus seems  neutral and the nuclear bar mode extinction could
 simply be due to classical Landau damping at \ILRs.

\begin{figure}
\centering
\includegraphics[keepaspectratio,width=\hsize]{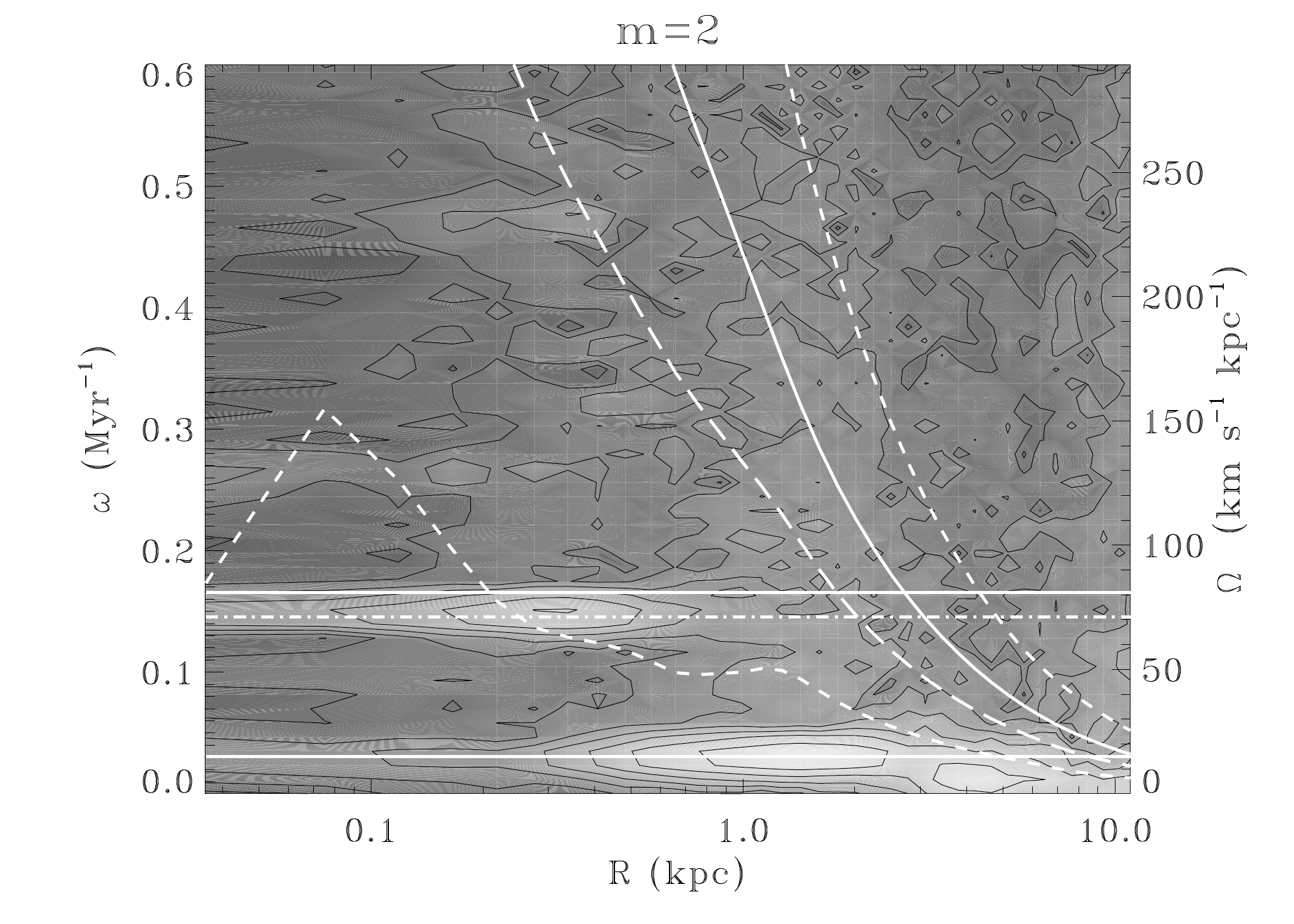}
\includegraphics[keepaspectratio,width=\hsize]{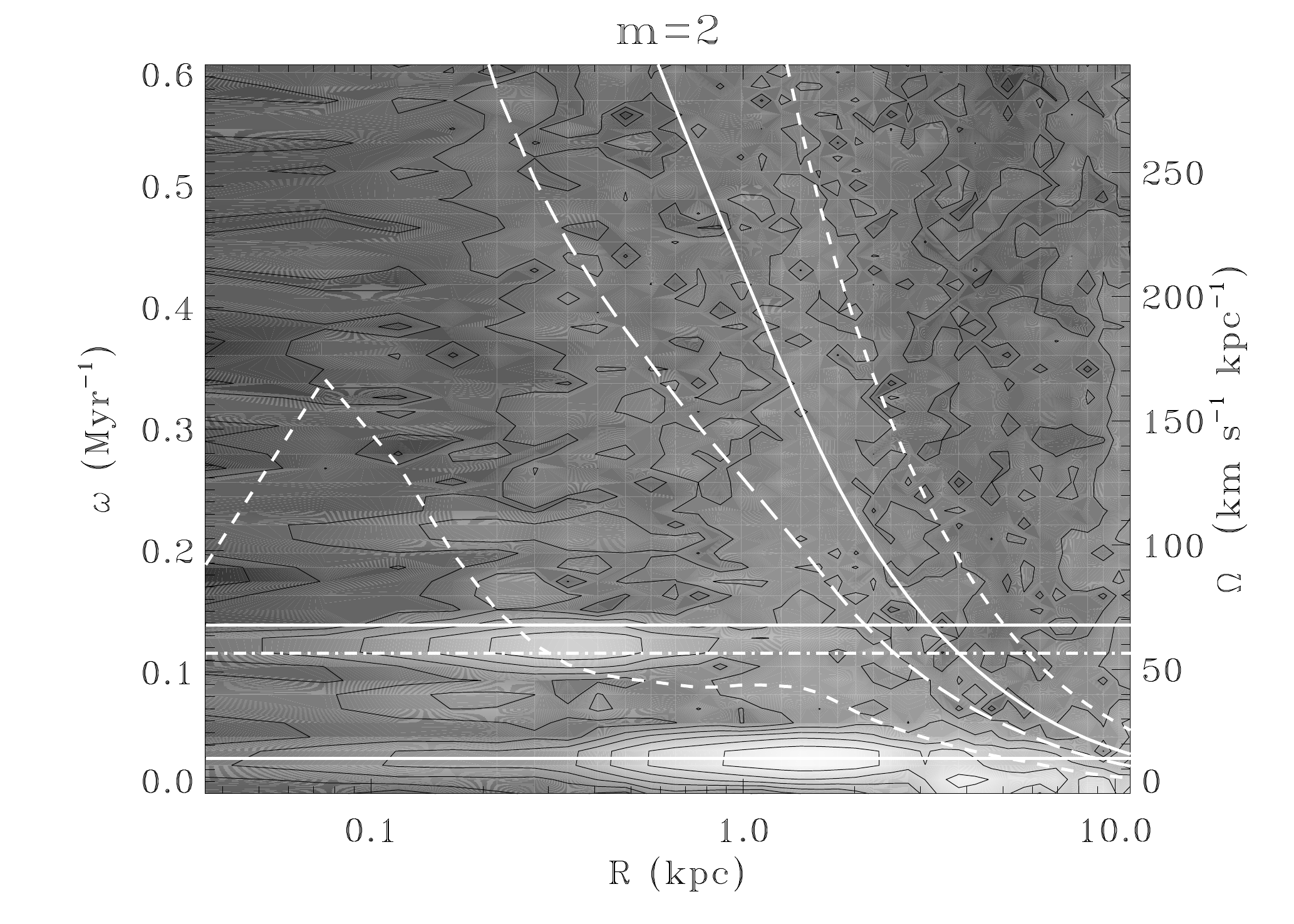}
\caption{$m=2$ power spectra for \ABAB\ centred at $t=7.6$ (top panel)
  just before \OLRs\ intersects \ILRp, $t=9.2$~Gyr (bottom) whereas
  the nuclear bar has been replaced by a large rotation oval. The time
  windows is 568~Myr wide.}
\label{fig:fft3b}
\end{figure}

In the case of \ABAB, the shape of the inner bar is oval for
$t\ga 7$~Gyr. Looking at the spectrograms (Fig.~\ref{fig:fft3b}) for
$t=7.6$ and 9.2~Gyr, it appears that the power of the inner bar mode
is concentrated just outside \ILRs\ whereas the maximum power of the
primary bar is just inside \ILRp, as it is for \FBFC. Again, it is hard to
find any mode between \CRs\ and \OLRs. Unlike \FBFC, several modes can
be detected at low level with $\Omega_p < \Omega < \Omega_s$ or
$\Omega > \Omega_s$, but only inside a region limited by
$\Omega(r)$. As we do for \FBFC, we can suspect Landau damping at \ILRs\ to
be responsible for the inner bar mode decrease.

\subsection{What  makes double-bars long-lived?}
\label{ssec:mecanism}

The lack of mode coupling is a major difference with previous
simulations. For pure N$-$body simulations, but also for models with
massless gas particles, this situation would prevent the nuclear bar
from being long-lived. Indeed, apart form the swing amplification at
corotation that does not operate here, there is no mechanism to bring
energy to the inner bar wave. Therefore, this wave should vanish by
Landau damping at \ILRs.

In our simulations, the first difference is the presence of a
self-gravitating gas component. This component behaves differently
from the stellar component because of its dissipative nature. However, as
shown for example by \citet{1993A&A...277...27F}, the gas does
not help to sustain a long-lived nuclear bar because the mass accumulation
close to the centre reinforces \ILRp\ , which in turn destabilizes the
nuclear bar orbits.

Another major difference is the star formation process that creates a
stellar population with initial dynamical properties inherited from
their parent gas elements (SPH particles). During the first Gyrs, when
the secondary bar grows, gas inflow and star formation are responsible
for bringing energy to the inner waves. When the central star
formation fades out as the mass of gas inflowing in the central region
decreases, the energy dissipation at \ILRs\ prevails. This is
basically the scenario of \FBFC\ for $t \ga 2$~Gyr. What delays the
inner bar extinction in \ABAB\ is a sustained local star formation
rate for several Gyr (cf. Fig.~\ref{fig:SFR}). In the same way as for sustaining the
$\sigma-$drop phenomenon \citep{2006MNRAS.369..853W}, the regular
feeding of the central region with recent stellar populations enables
secondary bars to be long-lived. Simulations by
\citet{1993A&A...277...27F} did not take star formation into account,
leading to the extinction of the secondary bar in roughly 5
turns. This is also the case of other past simulations with a gaseous
component that have been discussed before. A noticeable exception is
\citet{1996A&AS..118..461F} who ran a simulation with star formation,
but only on a short timescale (less than 2~Gyr).

\begin{figure}
\centering
\includegraphics[keepaspectratio,width=\hsize]{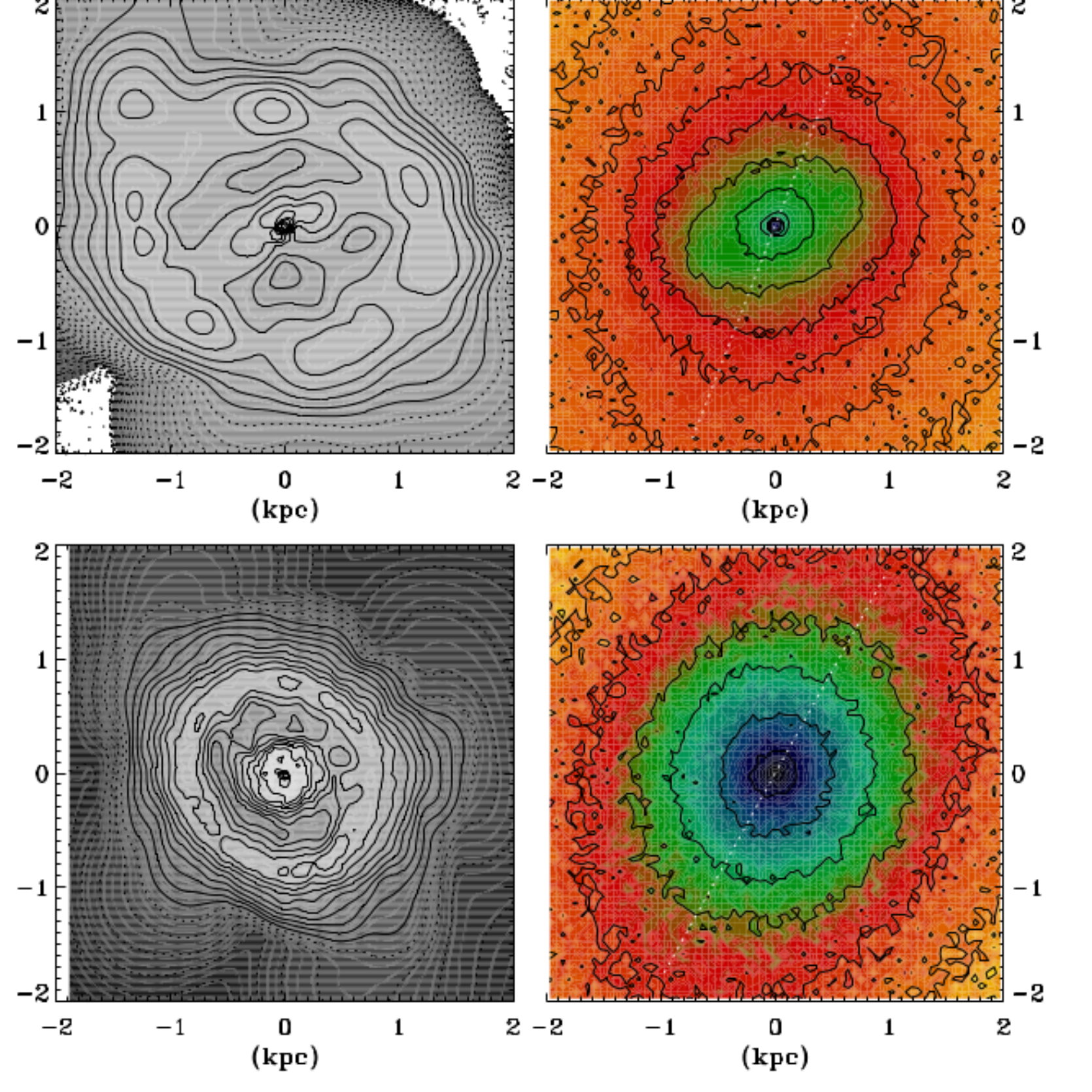}
\includegraphics[keepaspectratio,width=\hsize]{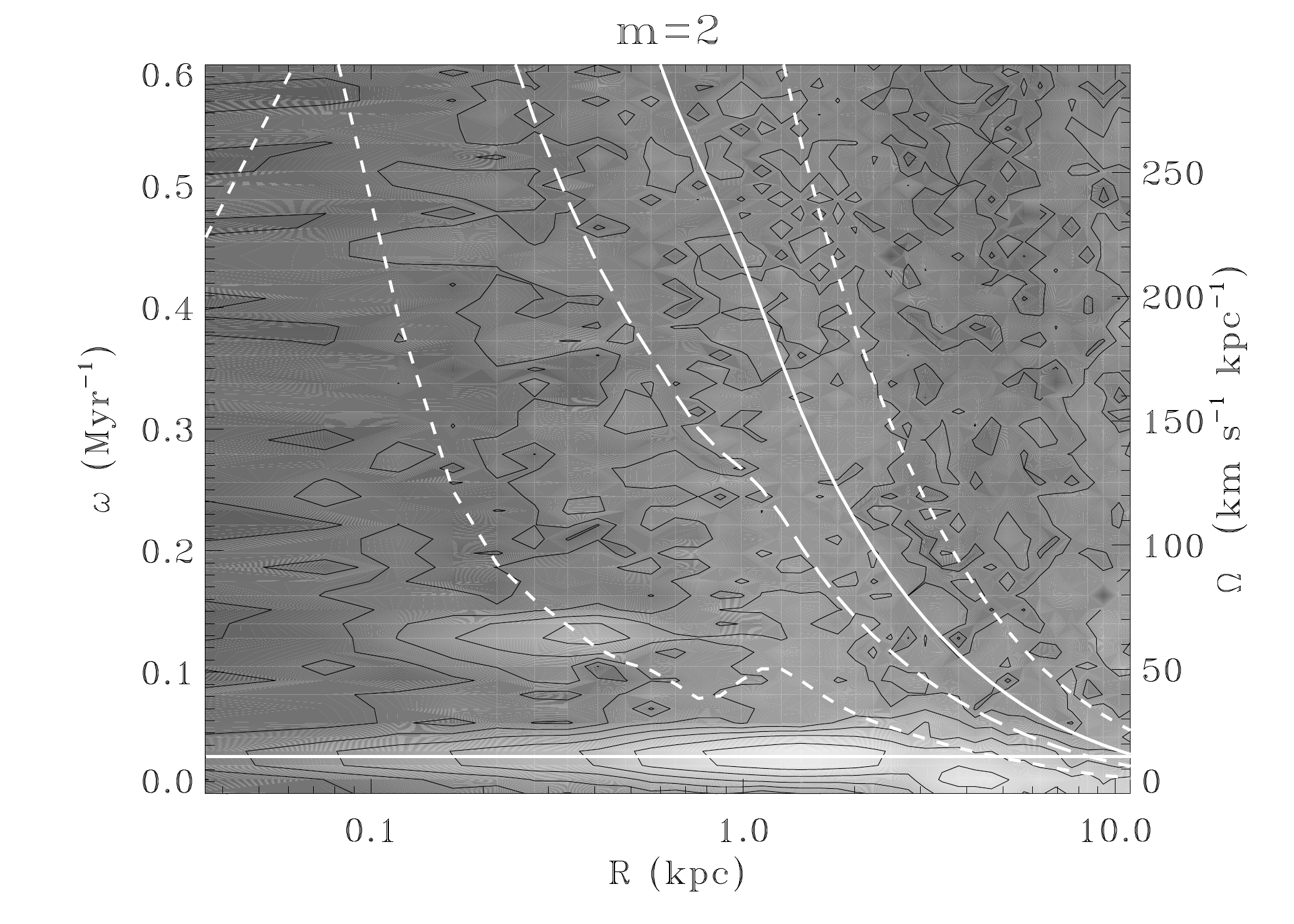}
\caption{Top and middle panels: as for Fig.~\ref{fig:abab} at
  $t=7.6$~Gyr, with (\ABAB, top) and without (\ABABnosf, middle) star
  formation. The surface brightness ranges from 22.8 to 17
  $\mathrm{Bmag}\,\mathrm{pc}^{-2}$ for \ABAB\ and 22.8 to 19.1
  $\mathrm{Bmag}\,\mathrm{pc}^{-2}$ for \ABABnosf. Bottom: $m=2$ power
  spectra for \ABABnosf, centred at $t=7.6$ (top panel). This figure
  can be directly compared with Fig.~\ref{fig:fft3b}. }
\label{fig:fftnosf}
\end{figure}

To firmly establish the role of star formation, we  recomputed
\ABAB\ from $t=5.25$~Gyr until $t=7.6$~Gyr switching off the star
formation process in the simulation code (\ABABnosf\ in short).
\citet{2006MNRAS.369..853W} used this method for
$\sigma-$drops. Figure~\ref{fig:fftnosf} shows the gas and stellar mass
distributions at $t=7.6$~Gyr. They are significantly different from
\ABAB.  The gas distribution is more concentrated in \ABABnosf\ than
in \ABAB\ since the gas is no longer consumed by the formation of new stars.
This has two immediate consequences. First, this predictably makes
dissolving  the inner bar easier as the gas can now reach the
centre, increasing there the mass that in turn axisymmetrizes the
gravitational potential. Second, no more gas is turned into stars so
that the mass of the inner stellar bar cannot increase. After roughly
2~Gyr, the inner bar is indeed barely detectable in the stellar
distribution of \ABABnosf. The spectrogram centred at $t=7.6$~Gyr, as
in Fig.~\ref{fig:fft3b}, confirms that, although it has not yet fully
disappeared, the amplitude of the $m=2$ mode has strongly decreased
with respect to \ABAB.

The role of star formation in sustaining the nuclear bar over several
Gyr is thus demonstrated for the first time.

\section{Conclusions}
\label{sec:conclusions}

Unlike the outcome of most numerical simulations with both stellar and
gaseous components, we have successfully simulated a \emph{long-lived}
inner bar embedded in a large-scale primary bar.

The ratio of the two bar lengths, the ratio of pattern speeds, as well as the age of
the inner stellar bar population and the central gas mass fit well
with observations published in the literature. Moreover, throughout the
simulation, our models go through various morphological phases
representative of the diversity of observations, including SB0.

The most important difference with past simulations leading to
short-lived double-barred galaxies is the lack of overlap between the
primary bar inner Lindblad resonance (\ILRp) and the nuclear bar
corotation (\CRs) and, more generally, the lack of any kind of resonance overlap.
The absence of mode coupling, confirmed by a Fourier analysis, implies
that to sustain a permanent nuclear bar mode another physical or
dynamical mechanism must feed the central waves.

Star formation in the central region is identified as possibly being 
responsible for bringing energy to the nuclear mode. Star formation is
also responsible for regulating the gas mass accumulation close to the
centre, in the sense that it prevents a strong increase in mass
density that can destabilize the inner bar orbits.

As a direct consequence (whereas the inner bar can be temporarily
undetectable, leading the pattern speed to be barely measurable or
unmeasurable), the corresponding perturbation modes survive and allow
the revival of the nuclear structure less than 1 Gyr after it
disappears, provided that star formation continues.

A side result of our study is that an overlap between the ILR of
the main bar and the CR of the nuclear bar cannot be systematically
assumed as it is not a necessary condition for the existence of
double-barred galaxies.
Another direct consequence of our simulations is the evidence that
several morphological features in barred galaxies that have been
identified by various names (double-bars, pseudo-bulges, triaxial
bulges, etc.)  originate dynamically from an unique inner
mode. \ABAB\ displays all these features at various stages of its
evolution. This speaks for a global analysis of triaxiality in the
central regions of barred galaxies as being a unique dynamical
phenomenon due to persistent modes in the central region.

\section{Acknowledgements}
I warmly thank Witold Maciejewski for fruitful discussions during the
meeting ``The Role of Bars in Galaxy Evolution'' in Granada, which
revived my long-lived interest in double-barred galaxies, and Isabel
P\'erez for organizing this productive meeting.  I am grateful to the
anonymous referee for his/her suggestions that have improved the
legibility of the paper. I  would also like to acknowledge the
\emph{P\^ole HPC} (High Performance Computing department) of the
University of Strasbourg for supporting this work by providing
technical support and access to computing resources. Part of the
computing resources were funded by the Equipex Equip@Meso
project. Local post-processing resources have been funded by the INSU
Programme National Cosmologie \&\ Galaxies.

\bibliography{AA_2014_25005}
\bibliographystyle{aa}

\end{document}